%% file: camera_ready.tex
\documentclass[acmtog, nonacm]{acmart}
\acmSubmissionID{130}

\usepackage{booktabs}
\usepackage[ruled]{algorithm2e}

\SetAlFnt{\small}
\SetAlCapFnt{\small}
\SetAlCapNameFnt{\small}
\SetAlCapHSkip{0pt}
\usepackage{multirow}
\usepackage{makecell}

\newcommand{\hl}[1]{{#1}}

\setcopyright{cc}
\setcctype{by-nc}
\acmJournal{TOG}
\acmYear{2025} 
\acmVolume{44} 
\acmNumber{4} 
\acmArticle{} 
\acmMonth{8} 
\acmPrice{}
\acmDOI{10.1145/3730822}

\begin{document}
\title{Improving Global Motion Estimation in Sparse IMU-based Motion Capture with Physics}

\author{Xinyu Yi}
\orcid{0000-0003-3504-3222}
\affiliation{%
  \institution{School of Software and BNRist, Tsinghua University}
  \city{Beijing}
  \country{China}}
\email{yixy20@mails.tsinghua.edu.cn}

\author{Shaohua Pan}
\orcid{0000-0002-6261-5268}
\affiliation{%
  \institution{School of Software and BNRist, Tsinghua University}
  \city{Beijing}
  \country{China}}
\email{isshpan@163.com}

\author{Feng Xu}
\orcid{0000-0002-0953-1057}
\affiliation{%
  \institution{School of Software and BNRist, Tsinghua University}
  \city{Beijing}
  \country{China}}
\email{xufeng2003@gmail.com}

\begin{abstract}
    \input{sec/0_abstract}  
\end{abstract}



\begin{teaserfigure}
  \includegraphics[width=\textwidth]{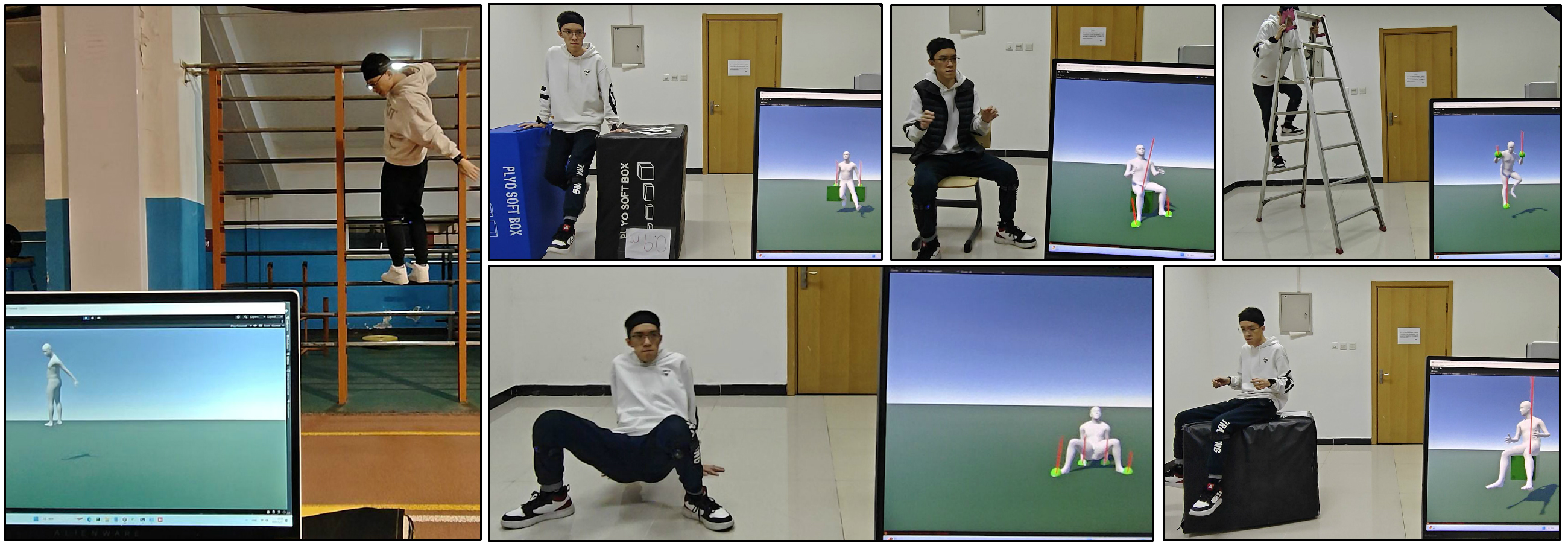}
  \caption{Live demos of our system showcasing unconstrained 3D-space motion capture (left). The method reconstructs human motion along with 3D-space contacts, contact forces, joint torques, and interacting proxy surfaces in real time (right).}
  \label{fig:teaser}
\end{teaserfigure}

\maketitle

\input{sec/1_intro}
\input{sec/2_related}
\input{sec/3_method}
\input{sec/4_experiments}
\input{sec/5_conclusion}

\begin{acks}
This work was supported by the National Key R\&D Program of China (2023YFC3305600), the Zhejiang Provincial Natural Science Foundation (LDT23F02024F02), and the NSFC (No.61822111, 62021002). This work was also supported by THUIBCS, Tsinghua University, and BLBCI, Beijing Municipal Education Commission. 
The authors would like to thank Wenbin Lin and Yunzhe Shao for their help on the live demos.
Feng Xu is the corresponding author.
\end{acks}

\bibliographystyle{ACM-Reference-Format}
\bibliography{cv}

\appendix
\input{sec/6_appendix}

\end{document}

%% file: sec/0_abstract.tex
By learning human motion priors, motion capture can be achieved by 6 inertial measurement units (IMUs) in recent years with the development of deep learning techniques, even though the sensor inputs are sparse and noisy.
However, human global motions are still challenging to be reconstructed by IMUs.  
This paper aims to solve this problem by involving physics.
It proposes a physical optimization scheme based on multiple contacts to enable physically plausible translation estimation in the full 3D space where the z-directional motion is usually challenging for previous works.
It also considers gravity in local pose estimation which well constrains human global orientations and refines local pose estimation in a joint estimation manner.
Experiments demonstrate that our method achieves more accurate motion capture for both local poses and global motions.
Furthermore, by deeply integrating physics, we can also estimate 3D contact, contact forces, joint torques, and interacting proxy surfaces.
Code is available at \url{https://xinyu-yi.github.io/GlobalPose/}.

%% file: sec/1_intro.tex
\section{Introduction}
Human motion capture, which focuses on digitalizing full-body human poses and movements, has long been studied and is crucial in various applications, such as augmented reality (AR), virtual reality (VR), gaming, robotics, and human-computer interaction (HCI).
%
%
Among the emerging methods, motion capture using sparsely worn inertial measurement units (IMUs) has gained attention due to its unique advantages. 
Unlike vision-based methods, IMU-based systems are not constrained by occlusions or the limitations of a fixed capture space, making them suitable for unconstrained environments and long-duration usage. 
Furthermore, the sparse setup significantly reduces cost compared to commercial systems like Xsens~\cite{Xsens}, which rely on dense IMU arrays that are expensive and intrusive. 
\par
Despite these advances, sparse IMU-based motion capture remains inadequate for real-world applications due to its relatively low accuracy, which comes from the strong noise in the raw signal of the IMU sensors, as mentioned by many previous works~\cite{DIP, yi2024pnp, TIP}.
For motion capture, as body movements always lie in the human pose space, the noise of the sensors can be reduced by constraining the final results in the prior space, which is the key reason that the noisy IMU sensors can still perform the motion capture task.
However, the local pose prior can do little on the global 6 degree of freedom (DOF) motion (containing global translation and orientation), which becomes a key challenge in IMU-based motion capture~\cite{EgoLocate}.
%
%
For global translation, some methods~\cite{PIP, yi2024pnp} use ground contacts to constrain human global translations to a 2D ground plane, meaning they cannot capture true 3D global movements like walking upstairs or lying on a bed.
For global orientation, most methods~\cite{DIP,PIP,yi2024pnp,TIP,zhang2024dynamic,van2024diffusionposer} directly rely on the IMU measurement of the root joint, leading to drift in long-duration tracking.
%
%
%
%
%
%
%
\par
%
%
Since the data-driven manner (learning local pose priors) cannot effectively denoise global motions, we propose to develop a physics-driven method to address this challenge.
We introduce a novel sparse IMU-based motion capture framework designed from the physics perspective, which enables unconstrained 3D-space translation estimation and improves global orientation accuracy, all while ensuring the physical correctness of the captured motion, benefiting both the global motion estimation and the local pose estimation.
Besides considering physics in motion capture, our method simultaneously estimates plausible physics-related information as byproducts, including 3D-space contacts, contact forces, joint torques, and interacting proxy surfaces, all from 6 IMUs, without being limited to the ground (see Fig.~\ref{fig:teaser}).
We believe the physical information extends the boundary of motion capture, making our technique more useful in topics like robotics and HCI.
\par
%
%
%
To refine global translation, most existing methods rely on ground contact estimation to involve stationary constraints, while 
%
%
%
\cite{TIP} takes a step further to estimate stationary points in 3D space in a data-driven manner. 
%
%
%
We argue that a joint data and physics-driven strategy is more powerful in handling this problem, and thus we further incorporate physics to estimate \textit{3D contacts} which provide additional information to constrain human bodies in 3D scenes. 
Our method selects a minimal set of stationary points that can physically explain human motion through contact forces, obtained by a physical optimization process.
%
By solving the 3D contacts in motion capture, besides the global translation, the physical plausibility of the estimated motion can also be benefited. 
%
%
%
%
\par
%
%
%

\begin{figure}
    \centering
    \includegraphics[width=0.9\linewidth]{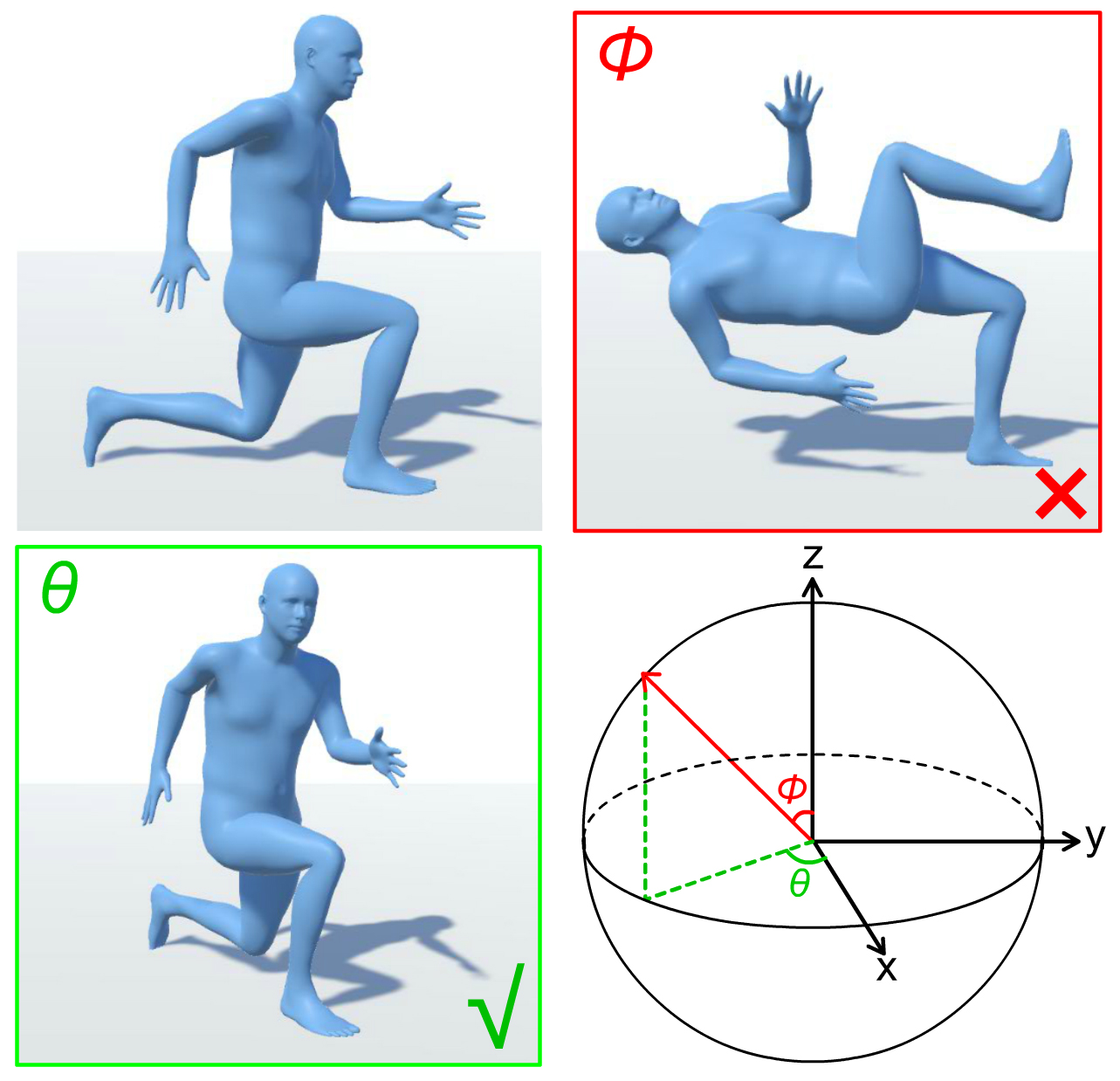}
    \caption{Illustration of the correlation between human local pose and global orientation. Given a local pose, the global $\phi$ orientation of the character is strongly constrained, while the heading direction $\theta$ can vary.}
    \label{fig:ill}
\end{figure}
Regarding global orientation, existing methods have not noticed the value of local pose information and thus rely solely on raw IMU measurements for the estimation.
%
%
However, by placing local poses within a physical coordinate system, such as a gravity-aligned frame, we observe that these local poses correlate with the global orientations represented in this system.
To better understand this correlation, consider a character in a specific local pose (Fig.~\ref{fig:ill}). 
While the character can have any $\theta$ value for its heading direction in a spherical coordinate system (gravity defined as the z-axis), its $\phi$ value is strongly constrained by the pose and, as a result, cannot be arbitrary.
%
%
%
%
Actually, the key here is that gravity influences human poses.
%
%
%
%
Based on this observation, we involve the gravity direction in local pose estimation by simultaneously reconstructing the local pose and refining the root-relative gravity direction.
With the refined gravity direction, we correct the global orientation error, as well as the errors in local poses. 
%
%
%
\par
In summary, our major contributions include:
\begin{itemize}
    \item A novel real-time motion capture system that captures world-aligned 3D human motion, along with 3D contacts, contact forces, joint torques, and interacting proxy surfaces, using only 6 IMUs.
    \item A joint data and physics-based 3D contact estimation method that enables unconstrained human translation estimation. 
    \item A gravity-aware pose estimation method that accurately estimates global orientations and local poses.
\end{itemize}

%% file: sec/2_related.tex
\section{Related Work}
\subsection{Human Motion Capture with Inertial Sensors}
We first review the works that capture human motion using wearable inertial sensors.
Commercial systems such as~\cite{Xsens} and~\cite{Noitom} provide high accuracy but rely on dense sensor setups, which are expensive and uncomfortable for everyday use. To reduce sensor count while maintaining accuracy, some research incorporates additional sensors to support the inertial sensors. For example, the works~\cite{hybridcap, robustcap, VIP, EgoLocate,lee2024mocap} use cameras to improve motion capture accuracy. However, the reliance on vision limits their application in certain scenarios where camera visibility may be obstructed or unavailable.
\cite{armani2024ultrainertialposer} utilizes UWB sensors to support IMU sensors,  but suffers from occlusion and requires careful calibration, restricting its use.
Other works~\cite{sparseposer, lobstr, ahuja2021coolmoves, aliakbarian2022flag, Aliakbarian_2023_ICCV, castillo2023bodiffusion, dittadi2021full, avatarposer, jiang2023egoposer, agrol, shin2023utilizing, neural3points, Zheng_2023_ICCV, questenvsim, questsim} employ 6DoF trackers, which provide both positional and orientation information to track human motion.
However, these trackers require external stations or cameras, limiting capture environments.
IMU-based methods, on the other hand, eliminate these limitations.
SIP \cite{SIP} reduces the number of IMUs to six using offline optimization techniques.
DIP \cite{DIP} leverages deep neural networks to estimate human pose in real time.
TransPose \cite{TransPose} uses multi-stage estimation and contact-foot-based fusion to reconstruct both human pose and translation.
PIP \cite{PIP} enhances TransPose further by integrating physics-based optimization with a flat-ground assumption.
TIP \cite{TIP} resolves pose ambiguity using stationary point estimation and reconstructs the height map of the capture environment.
DiffusionPoser \cite{van2024diffusionposer} inpaints the noise during the diffusion denoising process to support arbitrary IMU sensor configurations.
DynaIP \cite{zhang2024dynamic} incorporates additional real IMU data from Xsens datasets and regresses the body-part-based pseudo velocity, resulting in improved performance.
PNP \cite{yi2024pnp} models fictitious forces to fully utilize acceleration data for more accurate motion capture.
Some works~\cite{imuposer, xu2024mobileposer} further reduce the number of IMUs and utilize smart wearable devices to capture human motion.
However, none of these previous methods fully address global motion accuracy. Most of the existing approaches assume a flat ground to constrain global translation to a 2D plane and rely on noisy root IMU measurements to estimate global orientation. In contrast, our method integrates both data-driven and physics-based priors to enhance global motion estimation, leading to more accurate and physically plausible global motion reconstruction.
\subsection{Global Human Motion Estimation}
We review methods for estimating world-space global human motion, which includes both global orientation and translation.
Some approaches explore capturing world-space human motion using dynamic monocular cameras. GLAMR \cite{yuan2022glamr} predicts and optimizes human trajectories in the world coordinate system by infilling human motion sequences. SLAHMR \cite{ye2023slahmr} and PACE \cite{you2023pace} optimize both camera and human motion using SLAM results along with a learned human motion prior. WHAM \cite{shin2023wham} directly regresses global human motion in an autoregressive manner. WHAC \cite{yin2024whacworldgroundedhumanscameras} and TRAM \cite{wang2024tram} transform human motion from the camera coordinate to the world coordinate and refine the human trajectory. GVHMR \cite{shen2024gvhmr} predicts world-grounded human motion in a gravity-aware world coordinate.
Our method is similar to GVHMR in that it incorporates gravity information. However, unlike monocular-based motion capture methods, which transfer human motion into a gravity-aware frame, we transfer the gravity information into the human’s root frame and refine it during the estimation process. This approach enables heading direction invariance. For example, the same human local poses with different heading directions are considered as distinct poses in world coordinates. In contrast, when transferring gravity to the root frame, the gravity direction remains consistent.
\par
On the other hand, some works incorporate physics to improve global motion estimation, such as optimization-based methods \cite{tripathi2023ipman,Li2019, Rempe2020, PhysCap, Vondrak2012, Wei2010, Zell2017} and reinforcement-learning-based character control \cite{Bergamin2019, Isogawa2020, Liu2018, DeepMimic, SFV, Yu2021, Yuan2019, SimPoE, yao2024moconvq}.
For instance, PhysCap~\cite{PhysCap} employs physics-based motion optimization to adjust the global motion of the human, preventing unnatural leaning or depth jitter in monocular-based motion capture. 
\hl{IPMAN~\cite{tripathi2023ipman} predicts body pressure heatmaps and leverages intuitive physics to enforce physically plausible contacts, effectively mitigating unnatural floating and penetration artifacts in human reconstruction from color images. Recent works~\cite{MACS2024,zhang2024force} incorporate physical properties such as mass and friction into motion synthesis, enabling the generation of more natural and nuanced human body and hand-object interactions.}
In the context of IMU-based motion capture, works such as \cite{PIP, yi2024pnp} apply physics-based optimization to address issues like sliding, floating, and penetration, thus improving global translation accuracy.
However, \cite{PIP, yi2024pnp} assume a single flat ground to perform physics-based tracking. In contrast, our method enables 3D-space physics-based optimization by estimating 3D contacts and proxy surfaces directly from IMU measurements.

%% file: sec/3_method.tex
\begin{figure*}
    \centering
    \includegraphics[width=0.95\linewidth]{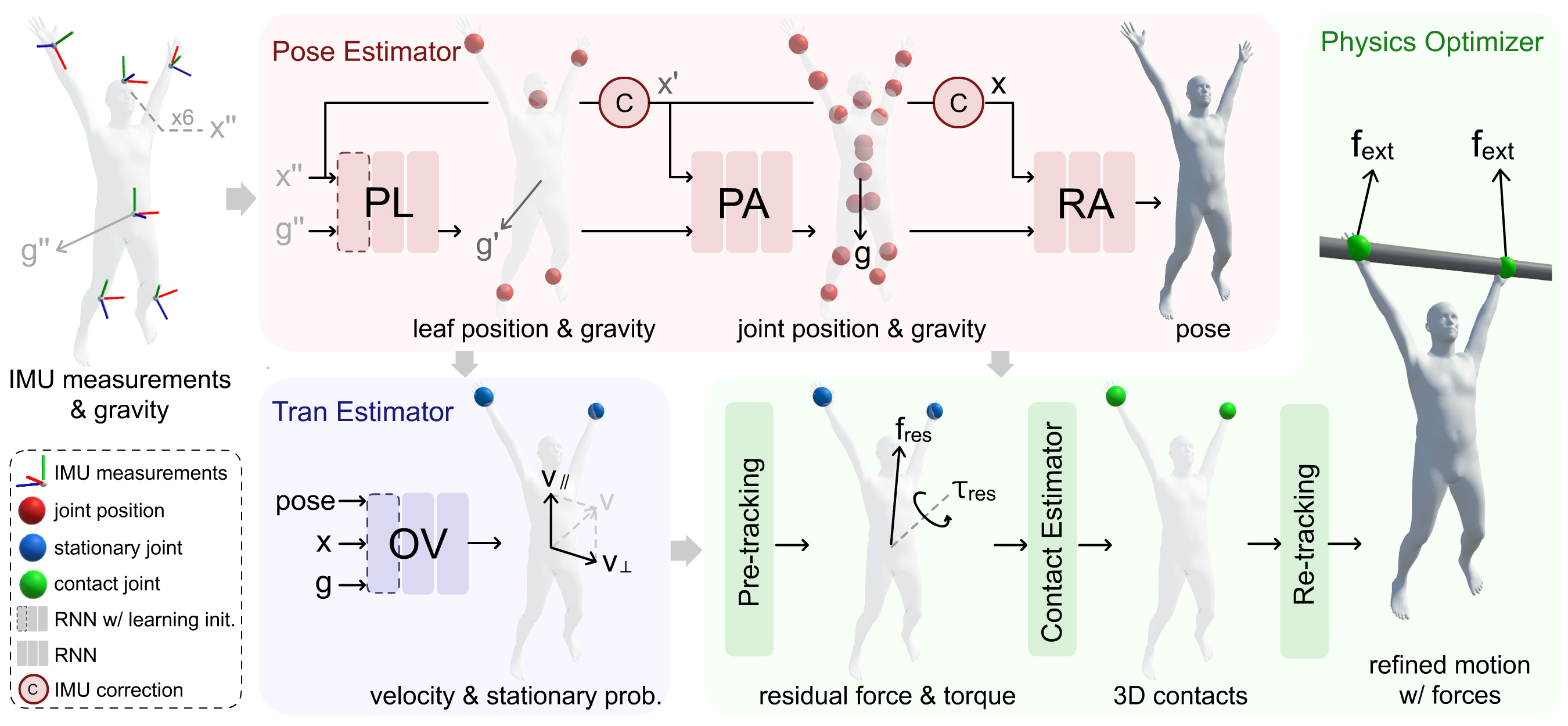}
    \caption{Overview of our method. We begin by estimating the human pose from IMU measurements (red). During this process, we simultaneously refine the root-relative gravity direction, which aids both local and global pose estimation. Next, we estimate human root velocity and joint stationary probability based on the pose and IMU measurements (blue). To incorporate gravity awareness, we decompose the root velocity into orthogonal components parallel and perpendicular to the gravity direction. Finally, we identify 3D contacts from the stationary joints using a physics-based algorithm, and perform physics optimization on the estimated motion (green).}
    \label{fig:method}
\end{figure*}

\section{Method}
Our task is to estimate human motion using 6 IMUs placed on the forearms, lower legs, head, and pelvis in real time.
Our system takes as input the inertial measurements of the sensors, including accelerations, angular velocities, and orientations. 
The output consists of human poses and 3D global motions, along with physical properties such as 3D-space contacts, contact forces, joint torques, and proxy surfaces with interactions.
\par
Our system consists of three modules: the \textit{pose estimator}, which estimates human pose (both local and global) from IMU measurements; the \textit{translation estimator}, which estimates global translation and stationary joints based on pose and IMU data; and the \textit{physics optimizer}, which identifies 3D contacts from the stationary joints and refines the human motion using physics-based optimization. 
The details of these modules are presented in Secs. \ref{sec:pose}, \ref{sec:tran}, and \ref{sec:phys}, respectively. 
Finally, we introduce our walking-based calibration method in Sec.~\ref{sec:calib}.
See Fig.~\ref{fig:method} for an overview of our method.
%
%
\subsection{Pose Estimator}\label{sec:pose}
The task of the pose estimator is to estimate human pose (defined as SMPL~\cite{SMPL} joint rotations) as well as the global orientation (defined as root rotation) from the IMU measurements. 
Our estimator is built upon PNP~\cite{yi2024pnp}, with the key improvement of integrating gravity information. 
We first discuss our advantages, followed by the details of the pose estimator framework.
\subsubsection{Gravity prior in pose estimation}
Estimating full-body pose from sparsely placed and noisy IMUs is inherently ambiguous.
Previous works \cite{DIP, TransPose, PIP, yi2024pnp, TIP, zhang2024dynamic} tackle this challenge by using deep neural networks to estimate local poses in the human root frame, aiming to model local pose priors to resolve the ambiguity. 
As a result, these methods are invariant to the global orientation of the human body.
However, we argue that there is a strong correlation between the human's local pose and the global orientation relative to a gravity-aligned world frame. 
For instance, a person lying flat is unlikely to perform walking poses. 
More precisely, the global orientation around the gravity axis is independent of the body's local pose (as a person can perform the same pose while facing different directions), while the orientations in the other degrees of freedoms (reflecting the body’s tilt) are correlated with the local pose (see Fig.~\ref{fig:ill}).
We find the root-frame gravity direction serves as a reliable indicator of these orientations, as it remains constant when the person rotates around the gravity axis but changes with variations in body tilt.
\par
Due to the correlation, we propose to model the \textit{joint prior distribution} of root-frame gravity direction and local pose by simultaneously reconstructing the local pose and refining the gravity. 
On one hand, an accurate gravity direction enhances local pose estimation by providing additional context beyond the root-relative inertial measurements. 
On the other hand, local pose estimation helps reduce the noise in the global orientation measured by the root's IMU.
By incorporating gravity prior into pose estimation, we introduce a more informative prior that improves the accuracy of both local and global pose estimation.
\subsubsection{Pose estimator pipeline}
Following PNP~\cite{yi2024pnp}, our pose estimation approach consists of three stages: first we estimate the leaf joint positions, then the full joint positions, and finally the human's pose.
\hl{This multi-stage design decomposes the complex task of pose estimation into simpler subtasks focused on joint position prediction and inverse kinematics, which has been shown to outperform single-stage pose estimation methods~\cite{TransPose}.}
Different from PNP, we additionally input the root-frame gravity direction at each stage and ask the network to simultaneously refine the gravity as an additional output. This enables the network to learn the joint prior distribution of local pose and global orientation.
Furthermore, we correct the IMU data based on the refined gravity at the beginning of each stage.
\par
The input to the pose estimator consists of \textit{1)} root-relative IMU measurements $\boldsymbol{x}''$, which is the concatenation of rotation matrices, angular velocities, and accelerations, and \textit{2)} the root-frame gravity direction $\boldsymbol{g}''$, which can be computed from the global orientation measured by the IMU on the root $\boldsymbol{R}''_{\mathrm{root}}$:
\begin{equation}
    \boldsymbol{g}'' = (\boldsymbol{R}''_{\mathrm{root}})^T\boldsymbol{g}_M,
\end{equation}
where $\boldsymbol{g}_M$ is the gravity direction in the world frame.
We use upper primes to indicate noise in the variables. Intuitively, the more primes a variable has, the noisier it is.
We begin by employing a Long Short-Term Memory (LSTM) network~\cite{LSTM}, denoted $PL$, to simultaneously reconstruct the root-relative leaf joint positions $\boldsymbol{p}_{\mathrm{leaf}}$ and refine the gravity vector to obtain $\boldsymbol{g}'$.
We then update the global orientation estimation by:
\begin{equation}\label{eq:imucorr1}
    \boldsymbol{R}'_{\mathrm{root}} = \boldsymbol{R}''_{\mathrm{root}}\mathbf{R}\left\{\boldsymbol{g}'\rightarrow\boldsymbol{g}''\right\},
\end{equation}
where $\boldsymbol{R}'_{\mathrm{root}}$ is the updated global orientation and $\mathbf{R}\left\{\boldsymbol{g}'\rightarrow\boldsymbol{g}''\right\}$ is the rotation matrix that rotates $\boldsymbol{g}'$ to $\boldsymbol{g}''$ with the minimal angle.
We then correct the root-relative IMU measurements based on the updated global root orientation, resulting in $\boldsymbol{x}'$. \hl{Note that the estimated leaf joint positions $\boldsymbol{p}_{\mathrm{leaf}}$ are relative to the ground-truth root frame. Thus, we do not need to update their values between stages. In contrast, the input IMU measurements $\boldsymbol{x}''$ are relative to the noisy root frame recorded by the IMU, requiring refinement across stages.}

Next, we concatenate $\boldsymbol{x}'$, $\boldsymbol{g}'$, and $\boldsymbol{p}_{\mathrm{leaf}}$, and feed this into a second LSTM network, $PA$, to estimate the full set of joint positions $\boldsymbol{p}_{\mathrm{all}}$ and further refine the gravity vector to $\boldsymbol{g}$. Once again, we correct the global orientation by:
\begin{equation}\label{eq:imucorr2}
    \boldsymbol{R}_{\mathrm{root}} = \boldsymbol{R}'_{\mathrm{root}}\mathbf{R}\left\{\boldsymbol{g}\rightarrow\boldsymbol{g}'\right\},
\end{equation}
and correspondingly update the root-relative IMU data to $\boldsymbol{x}$. \hl{The estimated joint positions $\boldsymbol{p}_{\mathrm{all}}$ are expressed in the ground-truth root frame and do not need refinement.}

Finally, we concatenate $\boldsymbol{x}$, $\boldsymbol{g}$, and $\boldsymbol{p}_{\mathrm{all}}$ and pass them through a third LSTM network, $RA$, to regress the body pose $\boldsymbol{\theta}$, using the 6D~\cite{6D} rotation representation.
The network structures and training details follow~\cite{yi2024pnp}.
The additional gravity vector is supervised using L2 loss.
Some small modifications are made to accelerate the training process, which are presented in App.~\ref{app:train}.

\hl{We would like to note that we denoise the root frame orientation by refining the root-relative gravity direction. This works because gravity is constant in the world frame but appears different when transformed into different root frames. By aligning the gravity, we effectively correct the root frame orientation. Specifically, we first input \(g''\) (relative to \(R''_{\mathrm{root}}\)) to estimate \(g'\) (relative to \(R'_{\mathrm{root}}\)), and then further refine it to obtain \(g\) (relative to \(R_{\mathrm{root}}\)), which serves as our final prediction of the root orientation. }

\subsection{Translation Estimator}\label{sec:tran}
The task of the translation estimator is to estimate the human's root velocity and stationary joints from pose and IMU data. 
We employ an LSTM network with the same structure as those in the pose estimator to perform this estimation.
To effectively incorporate gravity awareness, we input the root-frame gravity direction and ask the network to reconstruct two orthogonal components of the root joint’s velocity: one parallel to the gravity direction and the other perpendicular to it.
By doing so, we learn the conditional prior of global velocity conditioned on the body’s tilt represented by the root-frame gravity. 
Such approach is intuitive: a person lying down is unlikely to move as freely as a standing person, even if they share the same local pose.
On the other hand, the orthogonal velocities account for the fact that human translation in the gravity direction is often constrained and less free compared to movement in the horizontal plane.
\par
Specifically, we concatenate the following inputs: \textit{1)} the denoised IMU measurements $\boldsymbol x$, \textit{2)} the denoised gravity direction $\boldsymbol g$, \textit{3)} the estimated human pose $\boldsymbol{\theta}$, and \textit{4)} the joint positions computed by forward kinematics $\mathrm{FK}(\boldsymbol{\theta})$, all obtained from the pose estimator. 
We employ an LSTM network, denoted $OV$, to estimate the root velocity $\boldsymbol v$, specified by its two orthogonal components aligned with and perpendicular to the gravity direction $\boldsymbol v = \boldsymbol{v}_{\parallel} + \boldsymbol{v}_{\perp}$, along with the joint stationary probability $\boldsymbol s$.
\hl{For the gravity-aligned vector $\boldsymbol{v}_{\parallel}$, we predict only its magnitude, as its direction (i.e., gravity direction) is already estimated.}
For the joint stationary probability $\boldsymbol s$, we follow TIP~\cite{TIP} to consider five human joints for stationary estimation: the hands, the feet, and the pelvis, which results in a 5-D stationary probability.
After this estimation, we follow TransPose~\cite{TransPose} to use stationary constraints to refine the root velocity estimate:
\begin{equation}\label{eq:v}
    \min_{\tilde{\boldsymbol{v}}^{t}}{\|\tilde{\boldsymbol{v}}^{t}-\boldsymbol{v}^{t}\|^2+\sum_i{s_i\frac{1}{\Delta t^2}\|\mathrm{FK}_i(\boldsymbol{\theta}^{t},\tilde{\boldsymbol{v}}^{t}\Delta t)-\mathrm{FK}_i(\boldsymbol{\theta}^{t-1})\|^2}},
\end{equation}
where $\tilde{\boldsymbol{v}}^{t}$ is the refined root velocity, $\Delta t$ is the frame interval, the superscript $\cdot^t$ denotes the value at frame $t$, and the subscript $\cdot_i$ refers to the $i$-th joint.
This optimization tries to find a root velocity that keeps all stationary joints as fixed as possible, while making the root velocity as close as possible to the estimated value.
The analytical solution of Eq.~\ref{eq:v} is detailed in App.~\ref{app:infer}.

\subsection{Physics Optimizer}\label{sec:phys}
The task of the physics optimizer is to determine 3D contacts and perform physics-based optimization on the estimated motion.
Previous works~\cite{PIP, yi2024pnp} applied physics-based optimization under the assumption of flat ground.
This limitation prevents them from estimating 3D movements, such as walking upstairs.
For motions like sitting, as their system is unaware of the hip contact, a large virtual force (often called residual force) must be applied to the root joint to maintain balance, which is not physically correct (real humans do not have residual force on their root joint).
To address this, we extend their method to 3D space with a novel double-tracking approach.
We first provide an intuitive explanation of our method, followed by a detailed description of the optimizer.
\subsubsection{Method explanation}
The biggest challenge of performing physics-based optimization in 3D space is the lack of scene awareness, making contact detection impossible.
Thus, it is crucial to design a method that can estimate 3D contacts.
We observe that contact information can often be inferred from human motion: for example, if we see the motion of a person walking upwards, we naturally assume there are stairs beneath their feet.
This is based on how we understand the physical world.
We know that: \textit{1)} supporting forces are required to prevent the person from falling, and \textit{2)} the supporting force is more likely to act on the stationary foot.
By synthesizing this understanding, we design an algorithm that mimics this reasoning.
\par
We propose a double-tracking algorithm that determines 3D contacts and performs physics-based optimization on human motion.
First, we use a physical character to track the estimated motion without any contact.
To enable this tracking, we allow a large residual force~\cite{PhysCap} to act on the human root joint. 
This is akin to placing the physical character in an empty scene (without any object or ground) and allowing it to wear a rocket at the root joint that provides any external force needed.
These forces are not real (as real humans do not wear such a rocket) and must be explained by contact forces.
Thus, after pre-tracking, we select a minimal set of stationary joints that best explain the residual force through contact forces, yielding a set of 3D physical contacts.
Intuitively, this is like removing the rocket from the physical character and replacing it with forces applied to the selected contact joints.
Finally, using the identified contact joints and forces, we re-track the human motion to obtain the physically optimized result.
\par
We would like to note that this approach identifies contacts based on forces.
For instance, if a person lightly places their foot on a stair but keeps their body weight entirely on the grounded foot, the contact cannot be identified, as the motion can be explained by the supporting foot alone.
However, if the person shifts their weight onto the raised foot, our algorithm recognizes the foot as a necessary contact for maintaining balance.
\subsubsection{Phsyics-based optimization}
\paragraph{Physics model and notations} We use a torque-controlled floating-base character as the physics model.
This character shares the same skeleton structure and degrees of freedom as the SMPL~\cite{SMPL} model.
Its body mass, center of mass, and moment of inertia are extracted from the mean shape of the SMPL model, assuming a density of $1000\mathrm{kg/m^3}$.
In contrast to the SMPL model, which is driven by joint rotations, the physics character is driven by joint torques and external forces.
%
%
\par
Following the notations in PIP~\cite{PIP}, we denote the physics character's configuration (translation and pose) as $\boldsymbol{q}$, where the first three dimensions represent the global translation, followed by the root and other joints' rotations in local Euler angles. 
Joint torques are denoted as $\boldsymbol{\tau}$, which shares the same degree of freedom order as $\boldsymbol{q}$; specifically, the first 6 dimensions correspond to the residual forces and torques on the root joint, which should be zero in real humans~\cite{PIP}.
Global joint positions are represented by $\boldsymbol{r}$, and the global root position is represented by $\boldsymbol{p}$, which corresponds to the first three entries of $\boldsymbol{r}$.
Time derivatives are represented by adding a dot (for first derivatives) and double dots (for second derivatives) to the symbol (e.g., $\dot{\boldsymbol{r}}$ and $\ddot{\boldsymbol{r}}$ denote joint velocity and acceleration, respectively).
\par
When not considering contact forces, the character follows the equation of motion~\cite{RigidBodyDynamics} defined by:
\begin{equation}\label{eq:tau1} \boldsymbol{\tau} = \boldsymbol{M}(\boldsymbol{q})\ddot{\boldsymbol{q}} + \boldsymbol{h}(\boldsymbol{q}, \dot{\boldsymbol{q}}), \end{equation}
where $\boldsymbol{M}$ is the character's inertia matrix and $\boldsymbol{h}$ accounts for non-inertial effects and gravity.
This equation connects the physics character's torque $\boldsymbol{\tau}$ and acceleration $\ddot{\boldsymbol{q}}$ with the inertia $\boldsymbol{M}$.
Intuitively, this can be understood as analogous to Newton's second law: $\boldsymbol{F} = m\boldsymbol{a}$.
When considering the contact force, denoted as $\boldsymbol{\lambda}$, the equation of motion becomes:
\begin{equation}\label{eq:tau2}  \boldsymbol{\tau} + \boldsymbol{J}^T \boldsymbol{\lambda} = \boldsymbol{M}(\boldsymbol{q}) \ddot{\boldsymbol{q}} + \boldsymbol{h}(\boldsymbol{q}, \dot{\boldsymbol{q}}), \end{equation}
where $\boldsymbol{J}$ is the contact point Jacobian, which maps the force at the contact point to the torque on the character's degrees of freedom.
\hl{Analogously, the left side of Eq.~\ref{eq:tau2} computes the net force acting on the physical character, while the right side corresponds to the product of mass and acceleration.}
\paragraph{Pre-tracking} The goal of this stage is to estimate the forces required for the physics character to track the reference motion without incorporating contact forces.
First, we compute the reference joint rotations and positions for the tracking target.
The reference joint rotations are directly obtained from the pose estimator, denoted as $\boldsymbol{\theta}_{\mathrm{ref}}$ in the physics optimizer.
The reference joint positions $\boldsymbol{r}_{\mathrm{ref}}$ are computed as:
\begin{equation}\label{eq:r1} \tilde{\boldsymbol{r}}_{\mathrm{ref}} = \mathrm{FK}\left(\boldsymbol{\theta}_{\mathrm{ref}}, \boldsymbol{p} + \tilde{\boldsymbol{v}} \Delta t\right), \end{equation}
\begin{equation}\label{eq:r2} \boldsymbol{r}_{\mathrm{ref}} = \mathrm{Lerp}\left(\tilde{\boldsymbol{r}}_{\mathrm{ref}}, \boldsymbol{r}, \boldsymbol{s}\right), \end{equation}
where $\mathrm{Lerp}(a, b, t)$ is the linear interpolation function that interpolates between $a$ and $b$ by $t$. $\boldsymbol{p}$ and $\boldsymbol{r}$ are the current root and joint positions of the physics character, respectively, and $\boldsymbol{s}$ is the estimated joint stationary probability.
Eq.~\ref{eq:r1} computes the reference joint positions using forward kinematics on the estimated pose, with the updated root position. 
Eq.~\ref{eq:r2} further refines the reference joint positions by setting the stationary joint to be close to its current position.
\hl{Note that Eq.~\ref{eq:v} has already filtered error in global translation by optimizing root velocity to keep the stationary joint as stable as possible, considering all stationary joints globally. However, it does not modify the local pose. When there are multiple stationary joints, Eq.~\ref{eq:r2} further refines local pose by explicitly constraining individual joint positions, ensuring more precise stationary enforcement.}
\par
We then follow PIP~\cite{PIP} by employing dual PD controllers to compute the desired joint angular and linear accelerations, $\ddot{\boldsymbol{\theta}}_{\mathrm{des}}$ and $\ddot{\boldsymbol{r}}_{\mathrm{des}}$, that the physics character needs to generate in order to reproduce the reference motion:
\begin{equation}\label{eq:pd}
    \ddot{\boldsymbol{\theta}}_{\mathrm{des}}=k_{p_\theta}(\boldsymbol{\theta}_{\mathrm{ref}}-\boldsymbol{q}_{3:})-k_{d_\theta}\dot{\boldsymbol{q}}_{3:},
\end{equation}
\begin{equation}\label{eq:r3}
    \ddot{\boldsymbol{r}}_{\mathrm{des}}=k_{p_r}(\boldsymbol{r}_{\mathrm{ref}}-\boldsymbol{r})-k_{d_r}\dot{\boldsymbol{r}},
\end{equation}
where $k_{p_\theta}$, $k_{d_\theta}$, $k_{p_r}$, and $k_{d_r}$ are the gain parameters.
Intuitively, as long as the physics character generates the acceleration, it will follow the reference motion.
Next, we solve for the forces $\boldsymbol{\tau}$ required by the physics character to generate the desired accelerations:
\begin{equation}\label{eq:pretrack}
\begin{aligned}
    &\min_{\boldsymbol{\tau},\ddot{\boldsymbol{q}}}{\|\ddot{\boldsymbol{q}}_{3:}-\ddot{\boldsymbol{\theta}}_{\mathrm{des}}\|^2+\|\boldsymbol{J}\ddot{\boldsymbol{q}}+\dot{\boldsymbol{J}}\dot{\boldsymbol{q}}-\ddot{\boldsymbol{r}}_{\mathrm{des}}\|^2+\beta_\tau\|\boldsymbol{\tau}\|^2},\\
    &\quad\quad\quad\quad\quad\mathrm{s.t.}\,\,\,\,\boldsymbol{M}(\boldsymbol{q})\ddot{\boldsymbol{q}}+\boldsymbol{h}(\boldsymbol{q},\dot{\boldsymbol{q}})=\boldsymbol{\tau},
\end{aligned}
\end{equation}
where $\beta_\tau$ is used to control the relative weight of the regularization on forces.
The first two terms in Eq.~\ref{eq:pretrack} guide the physics character to generate the desired angular and linear accelerations respectively.
\hl{Specifically, \textit{1)} $\ddot{\boldsymbol{q}}_{3:}$ retrieves the angular acceleration of the physics character's joints, which should closely match the desired angular acceleration $\ddot{\boldsymbol{\theta}}_{\mathrm{des}}$; meanwhile, \textit{2)} $\ddot{\boldsymbol{r}}=\boldsymbol{J}\ddot{\boldsymbol{q}}+\dot{\boldsymbol{J}}\dot{\boldsymbol{q}}$ computes the linear acceleration of the physics character's joints, which should align with the desired linear acceleration $\ddot{\boldsymbol{r}}_{\mathrm{des}}$ from the dual PD controller.} 
The last term in Eq.~\ref{eq:pretrack} encourages the character to use relatively small forces to achieve the motion, preventing overshooting.
The equality constraint is the same as in Eq.~\ref{eq:tau1}, modeling the linear relationship between joint torques $\boldsymbol{\tau}$ and accelerations $\ddot{\boldsymbol{q}}$, without involving contact forces.
\hl{Intuitively, Eq.~\ref{eq:pretrack} finds a set of joint torques and forces that enable the physics character to replicate the reference motion.}
To accelerate the calculation, we reformulate Eq.~\ref{eq:pretrack} into an unconstrained sparse least squares problem, as detailed in App.~\ref{app:infer}.
Note that we are primarily concerned with the first 6 entries of the torque vector $\boldsymbol{\tau}_{:6}$, which correspond to the residual force and torque on the root joint that should be explained by physical contacts.
\paragraph{Contact estimation}
In this stage, we estimate the contact joints that explain the residual force and torque.
We begin by identifying contacts based on a set of rules.
A joint is marked as in contact if it is: \textit{1)} estimated to be stationary by the translation estimator, and in the meanwhile, \textit{2)} either in contact in the previous frame or currently touching the ground.
Additionally, if two stationary joints are sufficiently close to each other and at the same height, and one is judged to be in contact, we directly mark the other as in contact as well.
Any stationary joints not marked as contacts are labeled as potential contact joints.
\par
At this point, we have a set of contact joints and a set of potential contact joints. We then examine whether the current set of contact joints can support the human motion (i.e., explain the residual force). This is done by solving the following optimization problem:
\begin{equation}\label{eq:cont}
    \begin{aligned}
        &\min_{\boldsymbol{\lambda}}{\|(\boldsymbol{J}^T\boldsymbol{\lambda})_{:6}-\boldsymbol{\tau}_{:6}\|^2+\beta_\lambda\|\boldsymbol{\lambda}\|^2},\\
        &\quad\mathrm{s.t.}\quad \boldsymbol\lambda\in\text{friction cone},
    \end{aligned}
\end{equation}
where $\beta_\lambda$ is the regularization weight.
The first term in Eq.~\ref{eq:cont} tries to explain the residual force on the root joint by contact forces at the contact joints, while the second regularization term constrains the contact forces to remain small. The friction cone constraints ensure that the contact forces satisfy the Coulomb friction condition, as similar in \cite{PhysCap,PIP}.
Specifically, we apply friction cone constraints only on the forces exerted at the foot and pelvis joints, assuming the contact surface normal points toward the negative gravity direction, as these joints are typically in contact with horizontal surfaces. For the hand joints, except for the cases when the hands are touching the ground, we do not impose friction cone constraints, as the hands can grasp objects, allowing for arbitrary external forces.
By linearizing the friction cone constraints, Eq.~\ref{eq:cont} can be effectively solved using sparse quadratic programming, see~\cite{PhysCap}. 
After obtaining the optimal contact force $\boldsymbol{\lambda}$, we check the remaining residual force $\boldsymbol{e}$ that cannot be explained by the current set of contact joints:
\begin{equation}
    \boldsymbol{e}=\boldsymbol{\tau}_{:6}-(\boldsymbol{J}^T\boldsymbol{\lambda})_{:6}.
\end{equation}
If the magnitude of $\boldsymbol{e}$ exceeds a threshold $e_{\mathrm{th}}$, it indicates that the current set of contact joints cannot fully explain the residual force required by the character. 
In this case, we iteratively add potential contact joints to the set of contact joints, in increasing order of their distance to the ground, and redo the optimization in Eq.~\ref{eq:cont}.
If the residual force decreases by more than half when adding a potential contact joint, we accept it as a real contact. Otherwise, we reject it.
This process continues until either the remaining residual force falls below the threshold $e_{\mathrm{th}}$ or there are no more potential contact joints to add.
Finally, we obtain a set of contact joints and the corresponding contact force $\boldsymbol{\lambda}$ on these joints.
%
%

\paragraph{Re-tracking} With the estimated contact joints and forces, we re-track the reference motion in a more physically accurate manner. 
To prevent ground contacts from floating or penetrating the ground, we first update the reference positions of the ground contacts.
If a contact joint is above the ground within a distance threshold $d_{\mathrm{th}}$, we reduce its reference height by a factor of $0.1$, gradually pulling it towards the ground.
If a joint penetrates the ground, we update its reference position to the ground level.
Next, we recalculate the desired joint linear accelerations, denoted as $\ddot{\boldsymbol{r}}_{\mathrm{des}}^*$, following Eq.~\ref{eq:r3}, using the new reference joint positions. 
Finally, we perform the re-tracking defined by:
\begin{equation}\label{eq:retrack}
\begin{aligned}
    &\min_{\boldsymbol{\tau}^*,\ddot{\boldsymbol{q}}^*}{\|\ddot{\boldsymbol{q}}^*_{3:}-\ddot{\boldsymbol{\theta}}_{\mathrm{des}}\|^2+\|\boldsymbol{J}\ddot{\boldsymbol{q}}^*+\dot{\boldsymbol{J}}\dot{\boldsymbol{q}}-\ddot{\boldsymbol{r}}_{\mathrm{des}}^*\|^2+\beta_\tau^*\|\boldsymbol{\tau}^*\|^2},\\
    &\quad\quad\quad\mathrm{s.t.}\,\,\,\,\boldsymbol{M}(\boldsymbol{q})\ddot{\boldsymbol{q}}^*+\boldsymbol{h}(\boldsymbol{q},\dot{\boldsymbol{q}})=\boldsymbol{\tau}^*+\boldsymbol{J}^T\boldsymbol{\lambda}.
\end{aligned}
\end{equation}
Note that Eq.~\ref{eq:retrack} differs from Eq.~\ref{eq:pretrack} in that it incorporates the updated desired linear accelerations, the larger regularization weight $\beta_\tau^*$, and the equation of motion that accounts for the contact forces, which is introduced in Eq.~\ref{eq:tau2}. 
Intuitively, Eq.~\ref{eq:retrack} solves for the joint torques that, along with the external contact forces, enable the physics character to replicate the reference motion.
Solving this problem results in the joint torques used to drive the physics character, $\boldsymbol{\tau}^*$, and the resulting acceleration of the character, $\ddot{\boldsymbol{q}}^*$. 
We then update the character's state by:
\begin{equation}
    \boldsymbol{q}^{t} = \boldsymbol{q}^{t-1} + \dot{\boldsymbol{q}}^{t-1}\Delta t,
\end{equation}
\begin{equation}
    \dot{\boldsymbol{q}}^{t} = \dot{\boldsymbol{q}}^{t-1} + \ddot{\boldsymbol{q}}^*\Delta t.
\end{equation}
Finally, the refined translation and pose $\boldsymbol{q}^t$ are output.

\subsection{Walking Calibration}\label{sec:calib}
Inertial motion capture typically requires calibration due to variations in how users wear the IMUs~\cite{DIP,TransPose,TIP}. 
We propose a novel calibration method that simultaneously determines sensor-to-bone rotations and corrects sensors' relative drifts.
In previous methods~\cite{TransPose,PIP}, users need to take off all IMU sensors and place them with the same orientation to correct their relative drifts, and then put the sensors back on to perform a T-pose to determine sensor-to-bone rotations.
This two-step process is complex, time-consuming, and prone to errors due to any possible inaccurate placing or posing.
In contrast, our method only requires the user to take a single standard step forward, simultaneously correcting the drift and determining the rotations.
This is achieved by leveraging the prior knowledge of dynamic human walking motions, rather than static human poses only, where the integration of each IMU's acceleration is used to correct the sensors' relative drift and the known stepping poses help to calculate the sensor-to-bone rotations.
Our evaluations show that this novel calibration method contributes to better real-world applicability and performance.
\paragraph{Notations}
We denote the IMU sensor frame as $S$, the global inertial frame (the reference frame in which the IMU measures its orientation) as $I$, the SMPL~\cite{SMPL} bone frame as $B$, and the SMPL body-centric frame as $M$. 
The gravitational acceleration in the global inertial frame is denoted as $\boldsymbol{g}_I$.
IMU sensors typically measure raw acceleration and angular velocity in the sensor-local frame, denoted as $\boldsymbol{a}_S$ and $\boldsymbol{\omega}_S$, respectively. 
They also output the orientation with respect to the global inertial frame, denoted as $\boldsymbol{R}_{IS}$.
\par
The calibration process aims to determine the sensor-to-bone rotation $\boldsymbol{R}_{SB}$ for each sensor and the global extrinsic rotation $\boldsymbol{R}_{IM}$. For more details, readers are referred to  ~\cite{DIP, TransPose}. 
In our method, we also estimate the relative heading error $\boldsymbol{R}_1 \cdots \boldsymbol{R}_5$, which aligns the headings of the first five sensors with that of the sixth sensor. 
This process is typically known as "heading reset"~\cite{Xsens}, which traditionally requires aligning all sensors to the same orientation and resetting their yaw angle to the same value. 
However, our calibration method automatically corrects relative heading drift, without requiring the IMUs to be removed or realigned.
\paragraph{Walking-based calibration algorithm}
Our design takes inspiration from the commercial solutions~\cite{Xsens, Mocopi}, which also utilize a similar walking strategy.
We require the subject to stand straight first and record the IMU orientation measurements as $\boldsymbol{R}_{IS}^{(1)}$.
Next, the subject takes a step forward, and we integrate each sensor's global acceleration $\boldsymbol{a}_I$ during the step, computed from:
\begin{equation}
    \boldsymbol{a}_I=\boldsymbol{R}_{IS}\boldsymbol{a}_S+\boldsymbol{g}_I.
\end{equation}
The integration for each IMU is done by:
\begin{equation}
\begin{aligned}
    \boldsymbol{p}^{t+1}&=\boldsymbol{p}^{t} + \boldsymbol{v}^{t}\Delta t + 0.5\boldsymbol{a}^t\Delta t^2,\\
    \boldsymbol{v}^{t+1}&=\boldsymbol{v}^{t} + \boldsymbol{a}^{t}\Delta t,
\end{aligned}
\end{equation}
where $\boldsymbol{p}$ and $\boldsymbol{v}$ are the estimated position and velocity, respectively. The integration starts from zero position and velocity. Here and in the following, we omit the reference frame $I$ for simplicity.
To stabilize the integration, we apply the Zero-Velocity Update (ZUPT) technique~\cite{zupt}, which corrects the integration error by using the zero velocity observation when the subject finishes the step and stands still. 
Concretely, we track the variance of the velocity $\boldsymbol{v}$, denoted as $\sigma_{vv}$, and the covariance between position $\boldsymbol{p}$ and velocity $\boldsymbol{v}$, denoted as $\sigma_{pv}$, by:
\begin{equation}
\begin{aligned}
       \sigma_{pv}^{t+1} &= \sigma_{pv}^{t} + \sigma_{vv}^{t} \Delta t,\\
       \sigma_{vv}^{t+1} &= \sigma_{vv}^{t} + 1.
\end{aligned}
\end{equation}
Once the subject stops and stands still, the integrated velocity should ideally be zero (in practice, it is rarely zero due to noise in the IMU accelerations). Based on the zero-velocity observation, we update the posterior estimate of $\boldsymbol{p}$ as:
\begin{equation}\label{eq:kalman}
    \tilde{\boldsymbol{p}}=\boldsymbol{p}-\frac{\sigma_{pv}}{\sigma_{vv}}\boldsymbol{v},
\end{equation}
where $\tilde{\boldsymbol{p}}$ is the corrected position.
\hl{This equation exploits the positive correlation between velocity and position. To illustrate, consider the double integration of noisy acceleration during a step: if velocity is overestimated (e.g., final velocity $> 0$), the integrated position will also tend to be overestimated. This relationship enables position correction when the final velocity is known. For the mathematical foundations of Eq.~\ref{eq:kalman}, readers can refer to the Kalman Filter algorithm~\cite{kalman1960new}.}
Additionally, since we assume the subject steps on flat ground, the translation should be horizontal, meaning the position vector $\boldsymbol{p}$ must be orthogonal to the gravity vector $\boldsymbol{g}$.
By enforcing this condition, we further update the position estimate as:
\begin{equation}
    \bar{\boldsymbol{p}} = \tilde{\boldsymbol{p}} - \frac{\tilde{\boldsymbol{p}}\cdot \boldsymbol{g}}{\boldsymbol{g}\cdot\boldsymbol{g}}\boldsymbol{g}.
\end{equation}
Finally, after the subject returns to a straight pose, we record the IMU orientation measurements again as $\boldsymbol{R}_{IS}^{(2)}$. 
This step is not strictly required, but it can be used to verify whether  $\boldsymbol{R}_{IS}^{(2)}$ is close to $\boldsymbol{R}_{IS}^{(1)}$, indicating if the subject is standing in a correct pose. 
If a significant difference is found, we simply redo the calibration process.
\par
We now compute the calibration matrices.
If the IMUs are not subject to drift, we should obtain the same $\bar{\boldsymbol{p}}_1 \cdots \bar{\boldsymbol{p}}_6$ for the 6 IMUs, since the person maintains the same pose before and after the step.
However, IMUs are often affected by heading drift (e.g., caused by magnetic interference). This causes the trajectories to diverge.
In such case, the relative heading error can be computed by aligning the trajectories of the first five IMUs to the sixth IMU:
\begin{equation}
    \boldsymbol{R}_i = \mathbf{R}\{\bar{\boldsymbol{p}}_i\rightarrow\bar{\boldsymbol{p}}_6\},\quad i = 1\cdots5.
\end{equation}
With these alignment matrices, we modify the $i$-th IMU orientation measurement to $\bar{\boldsymbol{R}}_{IS}=\boldsymbol{R}_i\boldsymbol{R}_{IS}$, where $\boldsymbol{R}_6=\boldsymbol{I}$ (the sixth IMU's orientation does not require modifications).
The global extrinsic rotation $\boldsymbol{R}_{IM}$ can be computed by:
\begin{equation}
    \boldsymbol{R}_{IM} = \begin{bmatrix} \frac{\bar{\boldsymbol{p}}_6}{|\bar{\boldsymbol{p}}_6|} \times\frac{\boldsymbol{g}}{|\boldsymbol{g}|} & -\frac{\boldsymbol{g}}{|\boldsymbol{g}|} & \frac{\bar{\boldsymbol{p}}_6}{|\bar{\boldsymbol{p}}_6|} \end{bmatrix}.
\end{equation}
The sensor-to-bone rotation $\boldsymbol{R}_{SB}$ for each sensor can be computed by:
\begin{equation}
    \boldsymbol{R}_{SB} =  \left(\bar{\boldsymbol{R}}^{(1)}_{IS}\right)^T \boldsymbol{R}_{IM}  \boldsymbol{R}_{MB},
\end{equation}
where $\bar{\boldsymbol{R}}^{(1)}_{IS}=\boldsymbol{R}_i\boldsymbol{R}^{(1)}_{IS}$ is the heading-corrected recorded orientation, and $\boldsymbol{R}_{MB}$ is the known SMPL joint rotation in the standing pose.

%% file: sec/4_experiments.tex
\section{Experiments}
In this section, we first provide the implementation details (Sec.~\ref{sec:implementation}). 
Next, we compare our method with previous sparse-IMU-based motion capture approaches (Sec.~\ref{sec:comparisons}). 
We then evaluate the key components of our method (Sec.~\ref{sec:evaluations}). 
Finally, we discuss the limitations (Sec.~\ref{sec:limitations}). 
For additional results, please refer to our supplemental video.
\subsection{Implementation Details}\label{sec:implementation}
\paragraph{Networks and training}
Our pose estimator consists of three LSTM networks: $PL$, $PA$, and $RA$, and the translation estimator includes one LSTM network, $OV$, all following the architecture in~\cite{yi2024pnp}. 
Among these networks, $PL$ and $OV$ use a learning-based initialization scheme, following~\cite{PIP}. 
All networks are initially trained independently with their respective inputs and outputs for 100 epochs, after which we jointly train all four networks for 200 epochs.
During the joint training, we disable gradient propagation in the IMU correction module (Eq.~\ref{eq:imucorr1} and \ref{eq:imucorr2}), which helps stabilize the training. 
All estimations are supervised using L2 loss, except that the stationary probability is supervised with binary cross-entropy loss. 
The pose output in $RA$ is additionally supervised by applying forward kinematics and minimizing the L2 loss between the estimated and ground-truth joint positions.
Other training details follow~\cite{yi2024pnp}.
\paragraph{Hyperparameters in physics optimizer}
The frame interval $\Delta t$ is $1/60$ seconds.
The gain parameters in the dual PD controllers (Eq.~\ref{eq:pd} and \ref{eq:r3}) are set to $k_{p_\theta}=k_{p_r}=3600$ and $k_{d_\theta}=k_{d_r}=60$, based on the Taylor expansion results (see~\cite{PIP}).
The regularization on joint torque during pre-tracking (Eq.~\ref{eq:pretrack}) and re-tracking (Eq.~\ref{eq:retrack}) is set to $\beta_\tau=10^{-3}/M$ and $\beta_\tau^*=3\beta_\tau$, where $M=80\mathrm{kg}$ is the approximate weight of the physics character. This is used to align the unit of forces to the accelerations.
The regularization on contact forces during contact estimation (Eq.~\ref{eq:cont}) is set to $\beta_\lambda=0.4$, and the friction coefficient for modeling the Coulomb friction constraint is set to $0.7$.
The distance threshold in the re-tracking is set to $d_{\mathrm{th}}=0.15\mathrm{m}$.
During contact estimation, if a joint's stationary probability exceeds $0.7$, it is marked as stationary. The threshold for determining whether a joint is touching the ground or if two joints are at the same height is set to $0.05\mathrm{m}$.
The threshold for stopping the contact estimation process is set to $e_{\mathrm{th}}=400$.
Since residual forces and stationary joint estimations can be noisy, we employ a counter for each potential contact joint. The counter increases when the contact estimation algorithm identifies it as a contact.
Only when the counter reaches 5 (i.e., the physics optimization continues to consider a stationary joint as in contact for 83 milliseconds), is the joint changed to a true contact.
This approach significantly reduces false positives.
\hl{\paragraph{Initialization} Since IMUs lack global positioning signals, our method assumes the person starts at $(0,0,0)$ with their body touching the ground. The ground height is then initialized as the lowest joint's height in the first frame.}

\paragraph{Datasets}
The training datasets include \textit{1)} AMASS~\cite{AMASS}, which is a motion capture dataset and we synthesize IMU measurements using the method in~\cite{yi2024pnp}, and \textit{2)} DIP-IMU~\cite{DIP}, which contains motion and real IMU data. We follow previous works~\cite{DIP,yi2024pnp, PIP, TIP} to split the training and test sets for DIP-IMU. Following these works, we train our method on the large synthetic AMASS dataset and fine-tune it on the relatively smaller DIP-IMU dataset with real IMU measurements.

The test datasets include \textit{1)} TotalCapture~\cite{TotalCapture}, which has two different calibrations. Following~\cite{yi2024pnp}, we adopt both in our experiments, referred to as Official Calibration (with a larger IMU orientation error of about 12.1 degrees) and DIP Calibration (with a smaller IMU orientation error of about 8.6 degrees); \textit{2)} the DIP-IMU test split, which lacks translation ground truth and is used to evaluate pose estimation; \textit{3)} Xsens datasets, as used in~\cite{zhang2024dynamic}, including AnDy~\cite{andy}, CIP~\cite{cip}, and UNIPD~\cite{guidolin2022unipd}. UNIPD contains minimal global movement, so we do not evaluate translation on this dataset; \hl{\textit{4)} Nymeria dataset~\cite{nymeria}, which is a large-scale multimodal dataset containing full-body motion and IMU data. Specifically, it consists of in-the-wild, long-duration (over 20 minutes) human motion sequences, which present significant challenges for sparse-IMU-based motion capture. We use this dataset to evaluate the robustness of our method in long-duration tracking scenarios.}
All test datasets include real IMU measurements.
\paragraph{Metrics}
To evaluate pose estimation, we adopt the following metrics: \textit{1)} SIP Error, the global orientation error of the hips and shoulders in degrees; \textit{2)} Angular Error, the global orientation error of all SMPL joints in degrees; \textit{3)} Positional Error, the 3D position error of all SMPL joints in centimeters; and \textit{4)} Mesh Error, the vertex error of the posed SMPL meshes.
We evaluate the pose using two different settings: \textit{1)} the local setting, where we align the estimated global root position and orientation with the ground truth before evaluating the pose metrics. This setting follows the same method used in previous works~\cite{TransPose,PIP,yi2024pnp}; and \textit{2)} the global setting, where only the global root position is aligned, and the full pose (including global orientation) is evaluated. This setting reflects the world-space pose accuracy, which is crucial for many applications.
For translation estimation, we plot the global translation error curve against the real traveled distance and report the average translation drift in percentage.
To assess physical accuracy, we evaluate: \textit{1)} Root Jitter, the time derivative of the global acceleration (i.e., jerk, reflecting the naturalness of the motion~\cite{Jerk}) of the root joint, in $10^3\mathrm{m/s^3}$; and \textit{2)} Joint Jitter, the average jerk of all joints, also in $10^3\mathrm{m/s^3}$.
In all these metrics, lower values indicate better performance.
\paragraph{Hardware and performance}
Our method can run in real-time at 120 fps on a computer equipped with an Intel(R) Core(TM) i9-13900KF CPU and an NVIDIA GeForce RTX 4090 Graphics card.
For the live demo, we use Noitom Lab sensors~\cite{Noitom}, which send inertial measurements at 60 fps, and thus the live demo runs at the same framerate.
The method is implemented in PyTorch~\cite{Pytorch}, and we develop a physics-based optimization library in C++, which implements key algorithms such as the Recursive Newton-Euler Method~\cite{RigidBodyDynamics} tailored for humans.

\begin{table*}[t]
\centering
\caption{%
    Comparisons on pose estimation with previous works. The local metrics evaluate the local (root-relative) pose, while the global metrics assess the full pose (including both the local pose and global orientation). The lowest errors and standard deviations are shown in bold respectively.
}
\resizebox{\linewidth}{!}{
\begin{tabular}{cccccccccccc}
\toprule
\multirow{2}{*}{Method} & \multicolumn{4}{c}{Local} & & \multicolumn{4}{c}{Global} & \multirow{2}{*}{\makecell{Root\\Jitter}}  
& \multirow{2}{*}{\makecell{Joint\\Jitter}} \\ \cmidrule{2-5} \cmidrule{7-10}
           & SIP Error               & Ang Error               & Pos Error              & Mesh Error             & 
           & SIP Error               & Ang Error               & Pos Error              & Mesh Error             &               &               \\ 
           
\hline \multicolumn{12}{c}{TotalCapture (Official Calibration)} \\ \hline
DIP        & 18.73$\pm$12.22         & 17.57$\pm$9.86          & 9.47$\pm$5.87          & 11.33$\pm$6.76         & 
           & 20.08$\pm$12.52         & 18.48$\pm$10.04         & 10.53$\pm$6.12         & 12.42$\pm$6.96         & -             & -             \\
TransPose  & 18.12$\pm$9.02          & 14.91$\pm$5.90          & 7.10$\pm$3.92          & 8.09$\pm$4.31          & 
           & 17.72$\pm$9.23          & 13.94$\pm$5.72          & 7.27$\pm$4.16          & 8.32$\pm$4.46          & 1.77          & 1.95          \\
TIP        & 15.62$\pm$8.11          & 14.45$\pm$5.80          & 6.76$\pm$3.60          & 7.79$\pm$4.06          & 
           & 17.26$\pm$8.34          & 16.67$\pm$6.16          & 9.06$\pm$4.30          & 10.17$\pm$4.84         & 1.24          & 1.74          \\
PIP        & 14.52$\pm$7.60          & 13.85$\pm$5.67          & 6.22$\pm$3.46          & 7.21$\pm$3.94          & 
           & 14.11$\pm$7.74          & 13.18$\pm$5.64          & 6.61$\pm$3.70          & 7.63$\pm$4.08          & \textbf{0.12} & \textbf{0.21} \\
PNP        & 11.35$\pm$5.88          & 11.10$\pm$4.90          & 4.89$\pm$2.74          & 5.60$\pm$3.04          & 
           & 13.95$\pm$6.77          & 13.54$\pm$5.70          & 7.37$\pm$3.72          & 8.23$\pm$4.04          & 0.16          & 0.27          \\
DynaIP-X   & 25.92$\pm$11.11         & 16.89$\pm$6.74          & 8.39$\pm$4.95          & 9.63$\pm$5.44          & 
           & 24.60$\pm$11.01         & 15.75$\pm$6.35          & 8.28$\pm$4.81          & 9.82$\pm$5.17          & -             & -             \\
DynaIP-XD  & 26.82$\pm$10.70         & 16.99$\pm$6.61          & 8.19$\pm$4.79          & 9.44$\pm$5.32          & 
           & 25.75$\pm$10.97         & 15.76$\pm$6.31          & 8.09$\pm$4.75          & 9.42$\pm$5.09          & -             & -             \\
DynaIP-AD  & 26.12$\pm$9.80          & 16.71$\pm$6.39          & 7.60$\pm$4.55          & 8.76$\pm$5.06          & 
           & 25.43$\pm$9.90          & 15.60$\pm$6.09          & 7.72$\pm$4.55          & 8.96$\pm$4.92          & -             & -             \\
Ours       & \textbf{10.17$\pm$5.10} & \textbf{10.16$\pm$4.51} & \textbf{4.31$\pm$2.37} & \textbf{4.96$\pm$2.65} & 
           & \textbf{10.87$\pm$5.22} & \textbf{10.55$\pm$4.55} & \textbf{4.31$\pm$2.38} & \textbf{5.02$\pm$2.63} & 0.21          & 0.37          \\  

\hline \multicolumn{12}{c}{TotalCapture (DIP Calibration)} \\ \hline
DIP        & 18.62$\pm$12.40         & 17.22$\pm$10.04         & 9.42$\pm$5.89          & 11.22$\pm$6.79         & 
           & 19.61$\pm$12.96         & 17.78$\pm$10.43         & 9.67$\pm$6.05          & 11.36$\pm$6.88         & -             & -             \\
TransPose  & 16.60$\pm$8.80          & 12.90$\pm$6.14          & 6.56$\pm$3.92          & 7.43$\pm$4.33          & 
           & 16.88$\pm$9.23          & 12.76$\pm$6.27          & 6.68$\pm$4.04          & 7.45$\pm$4.39          & 1.65          & 1.88          \\
TIP        & 13.22$\pm$7.47          & 12.30$\pm$5.83          & 5.81$\pm$3.41          & 6.80$\pm$3.90          & 
           & 15.55$\pm$8.03          & 14.56$\pm$6.38          & 7.86$\pm$3.92          & 8.93$\pm$4.44          & 1.20          & 1.69          \\
PIP        & 12.93$\pm$7.12          & 12.04$\pm$5.80          & 5.61$\pm$3.35          & 6.51$\pm$3.84          & 
           & 13.35$\pm$7.56          & 12.11$\pm$6.06          & 5.80$\pm$3.50          & 6.61$\pm$3.92          & \textbf{0.11} & \textbf{0.20} \\
PNP        & 10.89$\pm$5.83          & 10.45$\pm$5.07          & 4.74$\pm$2.68          & 5.45$\pm$3.05          & 
           & 11.76$\pm$6.25          & 11.12$\pm$5.46          & 5.32$\pm$2.95          & 6.04$\pm$3.31          & 0.15          & 0.26          \\
DynaIP-X   & 24.59$\pm$10.38         & 14.85$\pm$6.81          & 7.42$\pm$4.76          & 8.54$\pm$5.22          & 
           & 24.87$\pm$10.77         & 14.54$\pm$6.85          & 7.38$\pm$4.75          & 8.36$\pm$5.11          & -             & -             \\
DynaIP-XD  & 26.22$\pm$10.40         & 15.11$\pm$6.80          & 7.46$\pm$4.67          & 8.66$\pm$5.15          & 
           & 26.55$\pm$11.02         & 14.81$\pm$6.88          & 7.46$\pm$4.68          & 8.50$\pm$5.04          & -             & -             \\
DynaIP-AD  & 27.20$\pm$10.27         & 15.17$\pm$6.78          & 7.52$\pm$4.66          & 8.54$\pm$5.16          & 
           & 27.43$\pm$10.73         & 14.95$\pm$6.85          & 7.60$\pm$4.72          & 8.52$\pm$5.12          & -             & -             \\
Ours       & \textbf{9.81$\pm$5.06}  & \textbf{9.99$\pm$4.78}  & \textbf{4.25$\pm$2.41} & \textbf{4.94$\pm$2.75} & 
           & \textbf{10.24$\pm$5.41} & \textbf{10.15$\pm$5.05} & \textbf{4.18$\pm$2.44} & \textbf{4.87$\pm$2.76} & 0.20          & 0.35          \\   

\hline \multicolumn{12}{c}{DIP-IMU} \\ \hline
DIP        & 17.35$\pm$9.56          & 15.36$\pm$8.55          & 7.59$\pm$4.19          & 9.05$\pm$4.93          & 
           & 17.33$\pm$9.54          & 15.41$\pm$8.59          & 7.61$\pm$4.18          & 9.05$\pm$4.90          & -             & -             \\
TransPose  & 17.06$\pm$8.95          & 8.86$\pm$4.82           & 6.03$\pm$3.72          & 7.17$\pm$4.29          & 
           & 16.98$\pm$8.90          & 8.76$\pm$4.75           & 6.00$\pm$3.68          & 7.12$\pm$4.24          & 0.95          & 1.11          \\
TIP        & 16.90$\pm$8.90          & 9.07$\pm$5.07           & 5.63$\pm$3.45          & 6.62$\pm$3.99          & 
           & 16.97$\pm$8.79          & 9.26$\pm$4.97           & 5.70$\pm$3.41          & 6.67$\pm$3.93          & 1.06          & 1.56          \\
PIP        & 15.33$\pm$7.89          & 8.78$\pm$4.75           & 5.12$\pm$3.05          & 6.02$\pm$3.56          & 
           & 15.30$\pm$7.75          & 8.99$\pm$4.77           & 5.27$\pm$3.06          & 6.13$\pm$3.55          & \textbf{0.11} & \textbf{0.17} \\
PNP        & 13.71$\pm$6.68          & 8.75$\pm$4.28           & 4.97$\pm$2.72          & 5.77$\pm$3.17          & 
           & 13.77$\pm$6.58          & 8.99$\pm$4.31           & 5.13$\pm$2.75          & 5.93$\pm$3.19          & \textbf{0.11} & \textbf{0.17} \\
DynaIP-X   & 17.39$\pm$8.80          & 8.88$\pm$4.46           & 5.92$\pm$3.23          & 7.16$\pm$3.80          & 
           & 17.27$\pm$8.70          & 8.61$\pm$4.28           & 5.84$\pm$3.14          & 7.05$\pm$3.68          & -             & -             \\
DynaIP-XD  & 13.75$\pm$7.14          & \textbf{7.05$\pm$3.93}  & 4.97$\pm$2.85          & 5.98$\pm$3.42          & 
           & 13.64$\pm$7.02          & \textbf{6.78$\pm$3.76}  & 4.91$\pm$2.77          & 5.89$\pm$3.33          & -             & -             \\
DynaIP-AD  & 14.46$\pm$7.47          & 7.12$\pm$\textbf{3.93}  & 5.13$\pm$2.97          & 6.17$\pm$3.54          & 
           & 14.39$\pm$7.38          & 6.85$\pm$\textbf{3.76}  & 5.09$\pm$2.90          & 6.10$\pm$3.45          & -             & -             \\
Ours       & \textbf{13.55$\pm$6.51} & 8.47$\pm$4.09           & \textbf{4.65$\pm$2.53} & \textbf{5.41$\pm$2.92} & 
           & \textbf{13.41$\pm$6.33} & 8.29$\pm$3.96           & \textbf{4.55$\pm$2.39} & \textbf{5.27$\pm$2.77} & 0.16          & 0.26          \\   
\bottomrule
\end{tabular}\label{tab:posecmp}
}
\end{table*}

\begin{figure*}
    \centering
    \includegraphics[width=0.9\linewidth]{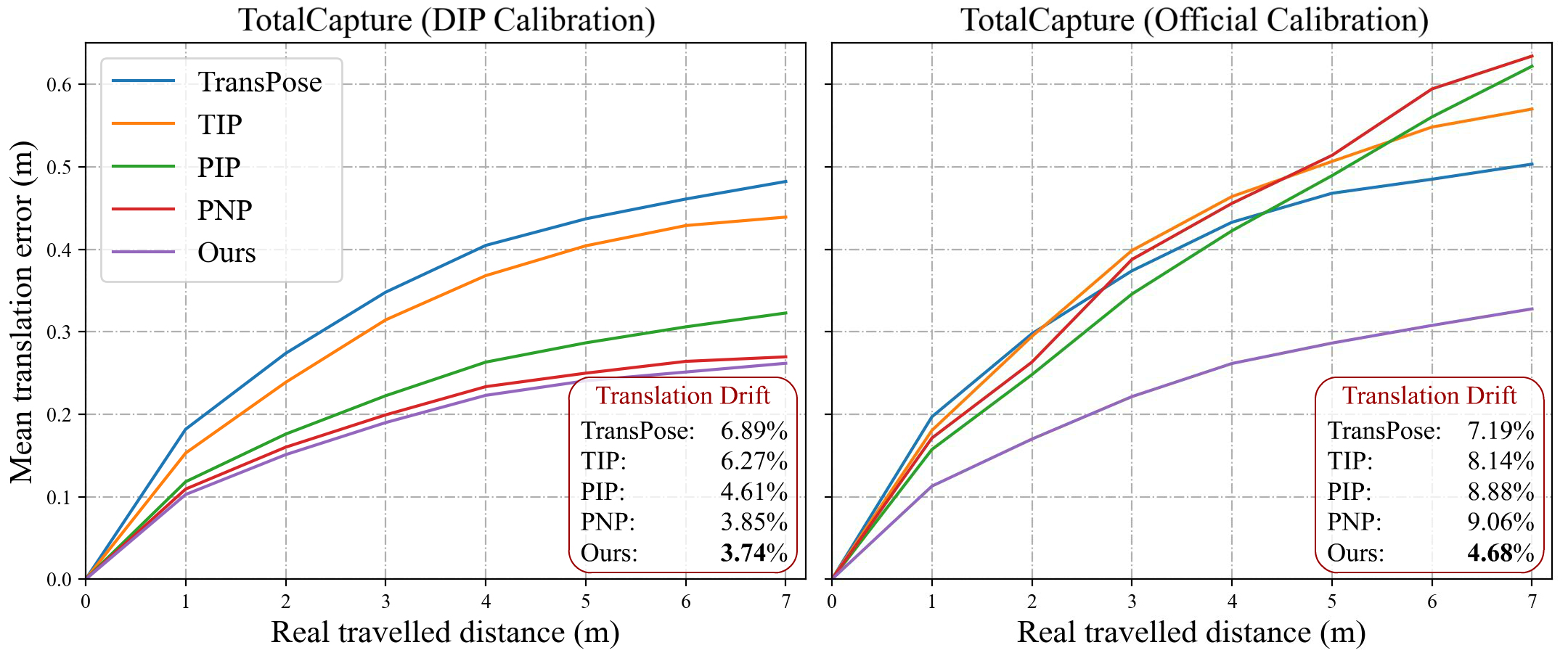}
    \caption{Translation comparisons on the TotalCapture dataset. We plot the cumulative translation error curves and report the average translation drifts at the 7-meter real travelled distance. A lower curve indicates better global translation accuracy.}
    \label{fig:trancmp}
\end{figure*}

\begin{table}[t]
\centering
\caption{%
    Additional comparisons on Xsens datasets. Ground-truth motions in these datasets are captured by Xsens~\cite{Xsens} and are transferred to the SMPL skeleton by~\cite{zhang2024dynamic}.
}
\resizebox{\linewidth}{!}{
\begin{tabular}{ccccccc}
\toprule
\multirow{2}{*}{Method} & \multicolumn{2}{c}{Local} & & \multicolumn{2}{c}{Global} & \multirow{2}{*}{\makecell{Trans\\Drift}} \\ \cmidrule{2-3} \cmidrule{5-6}
           & Pos Error              & Mesh Error             &  & Pos Error             & Mesh Error             &                 \\ 
       
\hline \multicolumn{7}{c}{AnDy} \\ \hline
PNP        & \textbf{5.84}$\pm$3.67 & 6.61$\pm$4.07          & & 5.75$\pm$3.33          & 6.27$\pm$3.45          & 4.04\%          \\
DynaIP-AD  & 6.62$\pm$5.15          & 7.75$\pm$6.02          & & 7.10$\pm$6.20          & 8.32$\pm$7.16          & -               \\
Ours       & 5.87$\pm$\textbf{3.38} & \textbf{6.47$\pm$3.65} & & \textbf{5.39$\pm$3.04} & \textbf{5.85$\pm$3.25} & \textbf{3.30}\% \\

\hline \multicolumn{7}{c}{CIP} \\ \hline
PNP        & 6.88$\pm$4.30          & 7.90$\pm$4.93          & & 7.11$\pm$4.45          & 8.13$\pm$5.10          & 5.63\%          \\
DynaIP-AD  & 6.30$\pm$4.17          & 7.10$\pm$4.47          & & 6.36$\pm$4.19          & 7.14$\pm$4.47          & -               \\
Ours       & \textbf{6.05$\pm$3.61} & \textbf{7.00$\pm$4.10} & & \textbf{5.60$\pm$3.18} & \textbf{6.40$\pm$3.58} & \textbf{4.80}\% \\

\hline \multicolumn{7}{c}{UNIPD} \\ \hline
PNP        & 4.11$\pm$2.40          & 4.65$\pm$2.69          & & 4.17$\pm$2.47          & 4.69$\pm$2.73           & -               \\
DynaIP-AD  & 4.25$\pm$2.45          & 4.62$\pm$2.57          & & 4.30$\pm$2.47          & 4.65$\pm$2.57           & -               \\
Ours       & \textbf{3.81$\pm$2.13} & \textbf{4.26$\pm$2.27} & & \textbf{3.73$\pm$1.99} & \textbf{4.10$\pm$2.07}  & -               \\
\bottomrule
\end{tabular}\label{tab:posecmp2}
}
\end{table}

\begin{table}[t]
\centering
\caption{%
\hl{Additional comparisons on Nymeria dataset. Inertial measurements are obtained from Xsens sensors using the method~\cite{zhang2024dynamic}.}}
\resizebox{\linewidth}{!}{
\begin{tabular}{cccccc}
\toprule
\multirow{2}{*}{Method} & \multicolumn{2}{c}{Local} & & \multicolumn{2}{c}{Global} \\ \cmidrule{2-3} \cmidrule{5-6}
           & Pos Error              & Mesh Error             &  & Pos Error             & Mesh Error                           \\ 
       
\hline \multicolumn{6}{c}{Nymeria} \\ \hline
PNP        & 7.87$\pm$4.01 & 8.81$\pm$4.46 & & 8.03$\pm$4.12 & 8.96$\pm$4.55 \\
Ours       & \textbf{7.25$\pm$3.46} & \textbf{8.28$\pm$3.87} & & \textbf{7.01$\pm$3.35} & \textbf{7.85$\pm$3.69} \\
\bottomrule
\end{tabular}\label{tab:posecmp3}
}
\end{table}


\begin{figure*}
    \centering
    \includegraphics[width=0.95\linewidth]{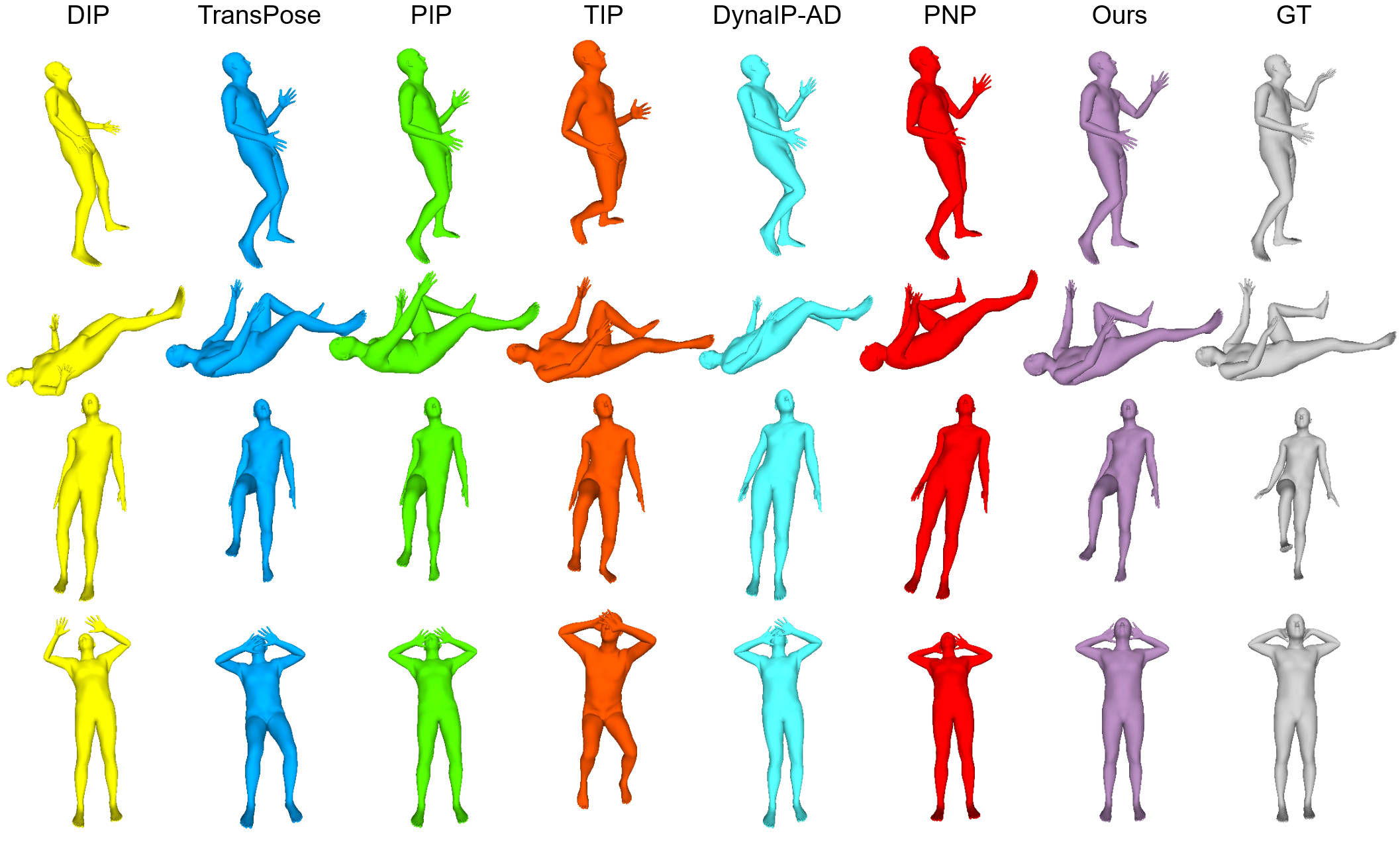}
    \caption{Qualitative comparisons on full pose estimation (including both local pose and global orientation). Results are picked from the TotalCapture dataset.}
    \label{fig:posecmp}
\end{figure*}

\begin{table*}[t]
\centering
\caption{%
    Ablation study on the pose estimator. We examine the effectiveness of incorporating gravity for local and full pose estimation.
}
\resizebox{\linewidth}{!}{
\begin{tabular}{cccccccccccc}
\toprule
\multirow{2}{*}{Method} & \multicolumn{4}{c}{Local} & & \multicolumn{4}{c}{Global} & \multirow{2}{*}{\makecell{Root\\Jitter}} & \multirow{2}{*}{\makecell{Joint\\Jitter}} \\ \cmidrule{2-5} \cmidrule{7-10}
& SIP Error & Ang Error & Pos Error & Mesh Error & & SIP Error & Ang Error & Pos Error & Mesh Error & & \\ 
           
\hline \multicolumn{12}{c}{TotalCapture (Official Calibration)} \\ \hline
w/o Grav Recon & 10.45$\pm$5.53 & 10.45$\pm$4.65 & 4.56$\pm$2.55 & 5.22$\pm$2.86 & & 12.66$\pm$5.71 & 12.57$\pm$4.72 & 6.59$\pm$2.95 & 7.42$\pm$3.26 & 0.34 & 0.56 \\
w/o Grav Input & 10.48$\pm$5.36 & 10.47$\pm$4.66 & 4.43$\pm$2.47 & 5.08$\pm$2.78 & & 12.54$\pm$5.57 & 12.40$\pm$4.73 & 6.48$\pm$2.87 & 7.27$\pm$3.17 & 0.33 & 0.55 \\
Ours           & \textbf{10.17$\pm$5.10} & \textbf{10.16$\pm$4.51} & \textbf{4.31$\pm$2.37} & \textbf{4.96$\pm$2.65} & 
               & \textbf{10.87$\pm$5.22} & \textbf{10.55$\pm$4.55} & \textbf{4.31$\pm$2.38} & \textbf{5.02$\pm$2.63} & \textbf{0.21} & \textbf{0.37} \\        

\hline \multicolumn{12}{c}{TotalCapture (DIP Calibration)} \\ \hline
w/o Grav Recon & 10.12$\pm$5.39 & 10.20$\pm$4.88 & 4.48$\pm$2.55 & 5.16$\pm$2.91 & & 10.88$\pm$5.68 & 10.72$\pm$5.13 & 4.93$\pm$2.68 & 5.63$\pm$3.00 & 0.30 & 0.50 \\
w/o Grav Input & 10.09$\pm$5.42 & 10.21$\pm$4.91 & 4.43$\pm$2.53 & 5.11$\pm$2.88 & & 10.94$\pm$5.73 & 10.78$\pm$5.20 & 4.88$\pm$2.65 & 5.57$\pm$2.97 & 0.30 & 0.50 \\
Ours           & \textbf{9.81$\pm$5.06}  & \textbf{9.99$\pm$4.78}  & \textbf{4.25$\pm$2.41} & \textbf{4.94$\pm$2.75} & 
               & \textbf{10.24$\pm$5.41} & \textbf{10.15$\pm$5.05} & \textbf{4.18$\pm$2.44} & \textbf{4.87$\pm$2.76} & \textbf{0.20} & \textbf{0.35} \\

\hline \multicolumn{12}{c}{DIP-IMU} \\ \hline
w/o Grav Recon & 15.01$\pm$6.98 & 9.33$\pm$4.45 & 5.05$\pm$2.70 & 5.83$\pm$3.09 & & 14.97$\pm$6.90 & 9.35$\pm$4.41 & 5.03$\pm$2.64 & 5.79$\pm$3.01 & 0.19 & 0.32 \\
w/o Grav Input & 14.02$\pm$6.73 & 8.86$\pm$4.33 & 4.98$\pm$2.71 & 5.73$\pm$3.12 & & 13.94$\pm$6.62 & 8.85$\pm$4.26 & 4.95$\pm$2.64 & 5.69$\pm$3.03 & 0.19 & 0.33 \\
Ours           & \textbf{13.55$\pm$6.51} & \textbf{8.47$\pm$4.09} & \textbf{4.65$\pm$2.53} & \textbf{5.41$\pm$2.92} & 
               & \textbf{13.41$\pm$6.33} & \textbf{8.29$\pm$3.96} & \textbf{4.55$\pm$2.39} & \textbf{5.27$\pm$2.77} & \textbf{0.16} & \textbf{0.26} \\     
\bottomrule
\end{tabular}\label{tab:abl1}
}
\end{table*}

\begin{table}[t]
\centering
\caption{%
    Ablation study on the pose estimator, the translation estimator, and the physics optimizer. We evaluate the global translation drift on the TotalCapture dataset with official calibration (OC) and DIP calibration (DC).
}
\resizebox{\linewidth}{!}{
\begin{tabular}{cccc}
\toprule
\multirow{2}{*}{Module} & \multirow{2}{*}{Method} & \multicolumn{2}{c}{Translation Drift} \\ \cmidrule{3-4}
                        &                         & TotalCapture (OC) & TotalCapture (DC) \\ \hline
\multirow{2}{*}{Pose}   & w/o Grav Recon          & 5.76\%            & 4.09\%            \\
                        & w/o Grav Input          & 5.55\%            & 3.90\%            \\ \hline

\multirow{2}{*}{Tran}  
                        & w/o Vel Decomp          & 5.14\%            & 3.79\%            \\
                        & w/o Grav Input          & 5.30\%            & \textbf{3.74}\%   \\ \hline

\multirow{2}{*}{Phys}   & w/o Physics             & 7.51\%            & 4.35\%            \\
                        & w/o Contact             & 6.09\%            & 4.36\%            \\ \hline

                        & Ours                    & \textbf{4.68}\%   & \textbf{3.74}\%   \\      
\bottomrule
\end{tabular}\label{tab:abl2}
}
\end{table}

\subsection{Comparisons}\label{sec:comparisons}
We compare our method with previous motion capture works that leverage sparse IMUs as input, including DIP~\cite{DIP}, TransPose~\cite{TransPose}, TIP~\cite{TIP}, PIP~\cite{PIP}, PNP~\cite{yi2024pnp}, and DynaIP~\cite{zhang2024dynamic}.
Specifically, we evaluate three versions of DynaIP: \textit{1)} DynaIP-X, which is trained on the Xsens datasets as described in~\cite{zhang2024dynamic}; \textit{2)} DynaIP-XD, which is trained on the Xsens datasets and then fine-tuned on the DIP-IMU train split; and \textit{3)} DynaIP-AD, which shares the same training datasets as the other methods, i.e., trained on the synthetic AMASS dataset and then fine-tuned on the DIP-IMU train split. The weights for this version are provided by the authors.
Note that DIP and DynaIP do not estimate global translations, so we do not include their results in the translation and jitter comparisons.
\par
We first compare our method with previous works on TotalCapture and DIP-IMU test split. The results are shown in Tab.~\ref{tab:posecmp}.
For pose estimation, our method consistently outperforms previous works in terms of both accuracy and standard deviation.
On one hand, our method achieves better local pose estimation accuracy, which can be attributed to incorporating gravity priors into the local pose estimation. The gravity direction provides additional physics-based information, aiding in the improvement of local pose regression.
On the other hand, the improvements in full pose (both global orientation and local pose) are significant, demonstrating the effectiveness of our method in reducing global orientation errors by refining the gravity direction. This is especially evident in the TotalCapture dataset with the official calibration, where the improvements in global pose are most notable due to the relatively large errors in the global orientation measurements.
While DynaIP achieves slightly better angular error on the DIP-IMU dataset, its generalization ability is weaker, as indicated by the significantly larger errors on the other two datasets.
In terms of jitter, our method achieves comparable results to PIP and PNP, and significantly outperforms works that do not incorporate physics, including TransPose and TIP.
\par
We then compare the translation estimation results on the TotalCapture dataset. The cumulative translation error is shown in Fig.~\ref{fig:trancmp}. Our method consistently achieves lower drift on the dataset under both calibration conditions.
It is important to note that this dataset was recorded on flat ground, and the works TransPose, PIP, and PNP all assume a flat ground, constraining human movements to the ground plane. In contrast, our method does not rely on this planar movement assumption, even achieving better translation accuracy.
Previous methods experience much larger drift under the official calibration, primarily due to the increased noise in the IMU measurements. By incorporating gravity and physics, our method effectively mitigates these noises, leading to significantly improved performance.
\par
We present additional pose and translation comparison results on the Xsens datasets, comparing with two state-of-the-art methods, PNP and DynaIP. The results are shown in Tab.~\ref{tab:posecmp2}. Our method demonstrates higher pose estimation accuracy and lower translation drift compared to previous works.
\hl{We also compare our method with PNP on the large-scale Nymeria dataset, and present the results in Tab.~\ref{tab:posecmp3}. This dataset features in-the-wild, long-duration tracking scenarios, making it particularly challenging for sparse-IMU-based motion capture. While our method exhibits slightly higher errors compared to the other test datasets, it still consistently outperforms previous work on estimation accuracy.} 
Rotational metrics are not included in these results as the datasets capture human surface rotation rather than bone rotation.
\par
Finally, we provide qualitative pose comparisons shown in Fig.~\ref{fig:posecmp}. The first row demonstrates that our method estimates the global orientation of the human more accurately, which can be attributed to the incorporation of gravity refinement. The second row highlights a more accurate local pose estimation by our method, showing that gravity awareness aids in local pose estimation.
For the more ambiguous cases in the last two rows, our method produces more accurate results, further emphasizing our advantage in deeply incorporating physics.

\begin{figure}
    \centering
    \includegraphics[width=0.835\linewidth]{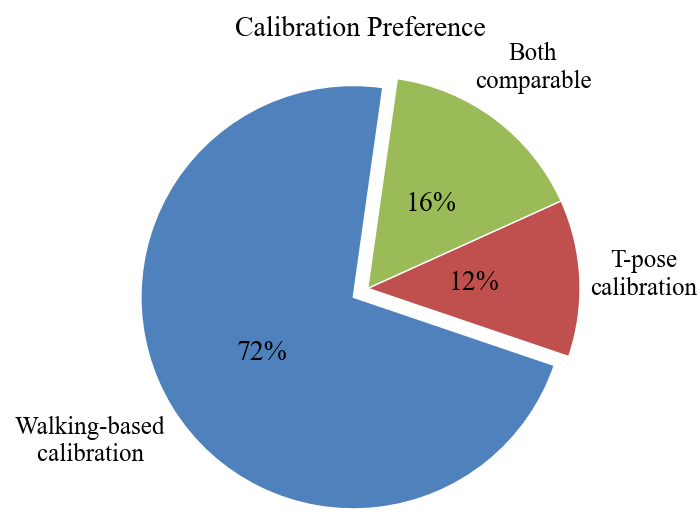}
    \caption{\hl{Voting results comparing walking-based calibration and T-pose calibration across 100 evaluations. Walking-based calibration was preferred in 72\% of cases, T-pose calibration in 12\%, and 16\% were rated as comparable.}}
    \label{fig:calib}
\end{figure}

\subsection{Evaluations}\label{sec:evaluations}
\paragraph{Evaluation on key modules}
In this section, we first conduct in-depth evaluations of the key components of our method.
For the pose estimator, we evaluate \textit{1)} \textit{w/o Grav Recon}, where we remove the gravity reconstruction in $PL$ and $PA$, and directly feed the noisy gravity direction obtained from the root IMU into the three stages; and \textit{2)} \textit{w/o Grav Input}, where we entirely remove the gravity direction from both the input and output of the three networks.
For the translation estimator, we evaluate \textit{1)} \textit{w/o Vel Decomp}, where we directly regress the root velocity using $OV$ without decomposing it w.r.t the gravity, and \textit{2)} \textit{w/o Grav Input}, where we remove the gravity input to $OV$ and also directly output the root velocity without decomposition. It should be noted that for the ablation study of the translation estimator, we only retrain the network $OV$ while keeping the weights of the pose estimator fixed.
For the physics optimizer, we evaluate \textit{1)} \textit{w/o Physics}, where we remove the entire physics optimizer and calculate the translation by integrating the estimated velocity; and \textit{2)} \textit{w/o Contact}, where we remove the contact estimation and the subsequent re-tracking stage, i.e., the physics character state is updated based on the pre-tracking results.
\par
We begin by evaluating the pose estimation using the TotalCapture and DIP-IMU datasets. Specifically, we investigate the necessity of incorporating gravity direction into the pose estimator. 
The results are presented in Tab.~\ref{tab:abl1}. 
Our full method demonstrates the best pose estimation accuracy and motion smoothness.
We also observe that merely inputting the gravity direction without reconstructing it often leads to worse results compared to not inputting it.
We attribute this to the fact that the raw gravity direction is usually too noisy for the network to effectively utilize. 
However, by refining the gravity though the networks, we achieve the best results for both local and full pose estimations.
\par
Furthermore, we evaluate the translation drift for the ablations of the three modules on the TotalCapture dataset. 
As depicted in Tab.~\ref{tab:abl2}, all the key components help reduce the global translation drift. 
Our method is particularly effective in the official-calibrated TotalCapture dataset, where the sensor is subject to larger noise. 
By integrating physics into the design of our method, we significantly mitigate such noise, leading to better global translation accuracy.

\hl{
\paragraph{Evaluation on calibration method} 
In this section, we demonstrate the advantages of our walking-based calibration method over the traditional T-pose calibration. Due to the lack of an existing dataset for quantitative comparison, we conducted a user study to evaluate the accuracy of our method. 
\par
In the study, five participants first performed a T-pose calibration, immediately followed by a walking-based calibration. They then executed a predefined 60-second motion while we recorded their IMU measurements and captured a reference video. Each participant repeated this process twice, resulting in 10 motion sequences. We processed these sequences using both calibration methods, producing 10 pairs of motion capture results for comparison. 
\par
Ten evaluators independently compared each pair against the reference video, choosing the better result or marking them as "comparable". The voting results are shown in Fig.~\ref{fig:calib}. Across the 100 evaluations, walking-based calibration was preferred in 72\% of cases, indicating that the walking-based calibration method generally produces more accurate motion capture results than T-pose calibration. We attribute this improvement to two main factors: \textit{1)} walking is easier for participants to perform accurately compared to holding a precise T-pose, and \textit{2)} walking-based calibration exploits human motion priors to mitigate relative sensor drift. For a qualitative comparison, readers are referred to the supplementary video.
\begin{table}[t]
\centering
\caption{\hl{Evaluation on long-term pose drift. We show global joint positional error (in cm) across three periods of a 20-minute outdoor sequence in the Nymeria dataset.}}
\begin{tabular}{cccc} 
\toprule 
& Period 1 & Period 2 & Period 3 \\ 
\midrule 
PNP & 7.30$\pm$4.23 & 7.41$\pm$4.23 & 8.23$\pm$6.72 \\ 
Ours & \textbf{6.18$\pm$3.42} & \textbf{6.52$\pm$3.58} & \textbf{6.38$\pm$3.37} \\ 
\bottomrule 
\end{tabular}\label{tab:drift}
\end{table}
\paragraph{Evaluation on long-term drift} 
To further evaluate long-duration tracking in unconstrained environments, we explored the Nymeria dataset and identified a 20-minute outdoor badminton sequence featuring fast, large-scale movements\footnote{The selected sequence name is 20231213\_s0\_shawn\_wright\_act5\_z5oir7. Readers can view this sequence on the official online data explorer at \url{https://explorer.projectaria.com/nymeria/20231213_s0_shawn_wright_act5_z5oir7?p=4&st=\%220\%22}.}.
We compared our method with previous state-of-the-art method PNP for real-time tracking of the entire 20-minute sequence and evaluated the global joint positional error across three evenly divided segments (0:00–6:35, 6:35–13:09, and 13:09–19:44, marked as Periods 1, 2, and 3) to assess potential drift over time. The results are shown in Tab.~\ref{tab:drift} (units in centimeters).
The results show no significant drift in our method, and we consistently outperform PNP in terms of pose accuracy. This stability is attributed to three key factors: \textit{1)} local pose drift is constrained by the learned pose prior from our gravity-involved pose estimation, \textit{2)} global pose drift is mitigated through gravity refinement, and \textit{3)} physics-based optimization further filters residual drift.
%
\par
To further visualize and compare the performance, we plot the smoothed global joint positional error curve for the entire sequence. As shown in Fig.~\ref{fig:drift}, our method consistently achieves lower errors across the sequence and shows no evident drift over time. Additionally, we select three representative frames from each of the three periods to qualitatively compare the pose reconstruction results of our method and PNP. Our method demonstrates significantly reduced drift, particularly in global orientation.
\begin{figure}
    \centering
    \includegraphics[width=\linewidth]{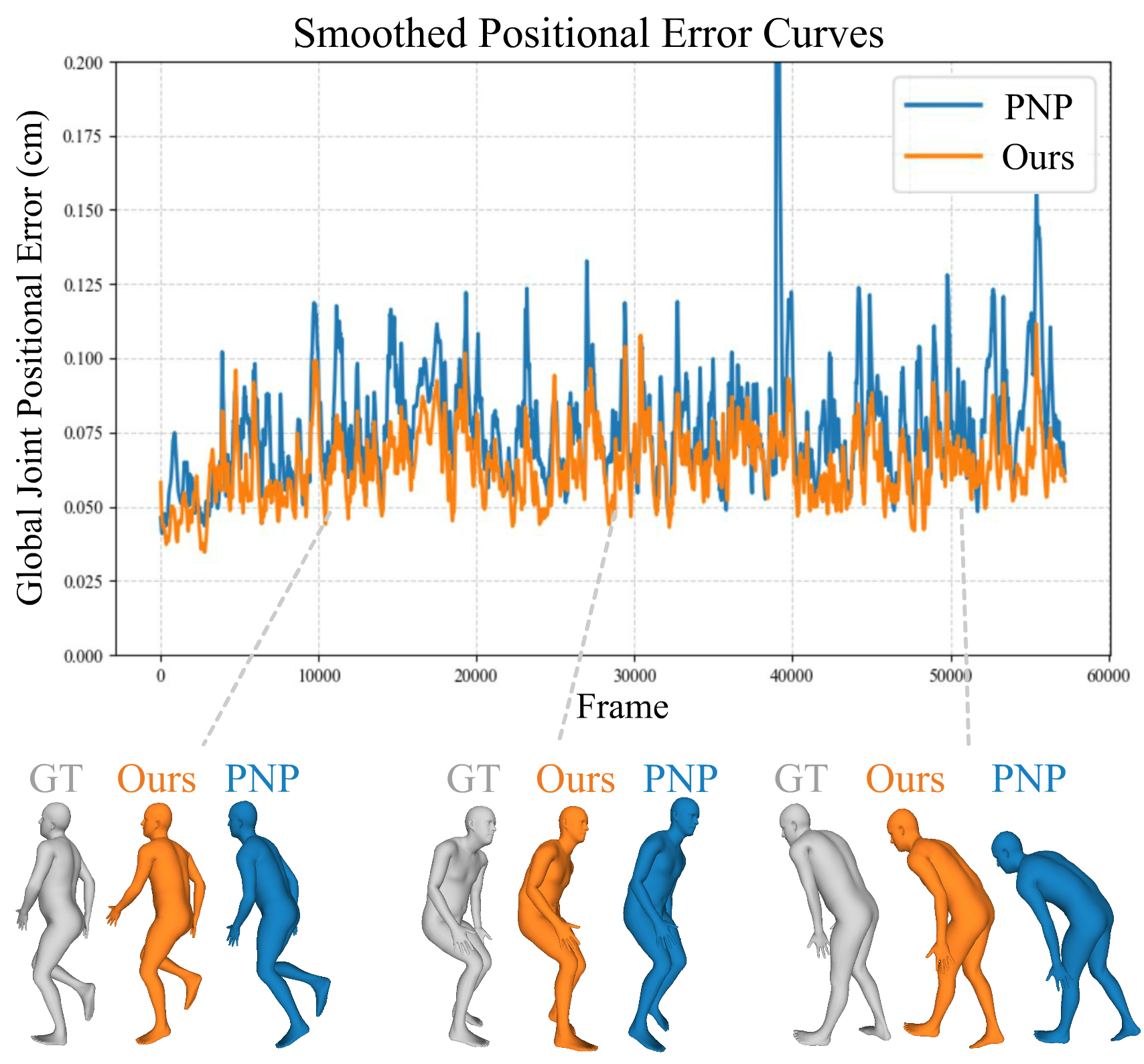}
    \caption{\hl{Smoothed global joint positional error curves over the 20-minute sequence. Our method consistently maintains lower error and shows no evident drift compared to PNP.}}
    \label{fig:drift}
\end{figure}
}

\begin{figure}
    \centering
    \includegraphics[width=0.9\linewidth]{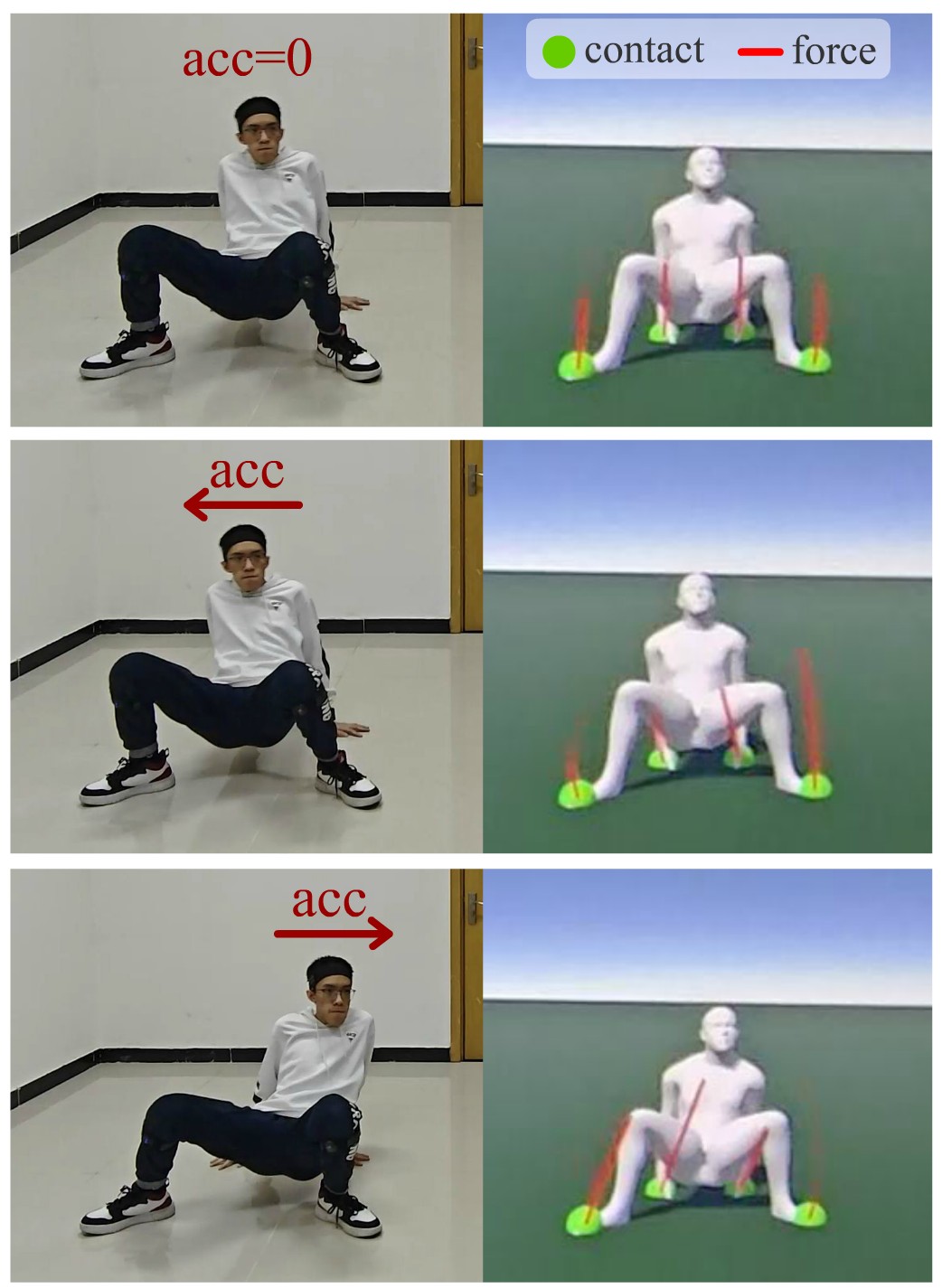}
    \caption{\hl{Visualization of contact force distribution during multi-contact motions. Forces are naturally adapted according to the body's acceleration, demonstrating that our method allows flexible and physically plausible force estimations.}}
    \label{fig:force}
\end{figure}

\subsection{Limitations}\label{sec:limitations}
\paragraph{Lack of 3D-space motion data} Our translation estimator is trained on the AMASS and DIP-IMU datasets, which contain limited 3D-space movements involving height changes (e.g., walking upstairs). This limitation affects the accuracy of vertical translation estimation. While our 3D contact estimation can help filter global translation estimates, incorporating more diverse motion data would further enhance translation accuracy.

\paragraph{Constrained contact joints} Our method estimates the stationary probability only for the hands, feet, and pelvis joints, restricting contact identification to these specific joints. Although this approach covers most scenarios, there are situations where other parts of the body may be in contact with objects, e.g., when leaning against a wall using the head.

\paragraph{Proxy surface and contact assumptions} Our method reconstructs stationary contacts based on forces, meaning sliding contacts or those with very slight forces cannot be accurately modeled. Additionally, we assume that proxy surfaces are horizontal when supporting the foot or pelvis, meaning tilted surfaces cannot be estimated. Finally, we cannot accurately capture very small height changes, such as walking onto a thin block, due to the estimation inaccuracies.

\hl{\paragraph{Ambiguous forces at multiple contacts} Resolving forces at multiple contacts is inherently ambiguous, as multiple solutions exist. Our method addresses this ambiguity through the regularization terms in Eq.~\ref{eq:pretrack} and \ref{eq:cont}, which minimize total human joint torque and contact force, respectively. Intuitively, among all solutions, this regularization encourages minimal forces and torques to reproduce the motion, aligning with the natural human tendency to minimize physical effort. Consequently, the regularization tends to distribute forces evenly across multiple contacts. Note that these regularization terms only serve as a soft constraint and do not dominate the optimization of Eq.~\ref{eq:pretrack} and \ref{eq:cont}. For instance, when the body moves during multi-contact movements, the contact force distribution adapts accordingly. Live demonstrations of such cases are shown in Fig.~\ref{fig:force}, and additional visualizations can be found in the supplementary video.
}

%% file: sec/5_conclusion.tex
\section{Conclusion}
In this paper, we propose a novel physics-driven approach to sparse IMU-based human motion capture, addressing key challenges in estimating global motion, specifically global translation and orientation. By integrating gravity priors into the framework, we significantly improve the accuracy of both local pose and global orientation estimation. Additionally, we enable unconstrained 3D-space motion estimation through physics-based 3D contact detection.
This combination of data-driven and physics-based priors results in more physically plausible motion capture, enhancing both realism and accuracy in real-world environments. Our method also produces valuable byproducts, including joint torques, contact forces, and interactions with proxy surfaces, expanding the potential applications of IMU-based motion capture.
Through extensive experiments, we show that our approach outperforms existing methods in both pose and translation accuracy, offering a robust, real-time, and cost-effective solution for motion capture in unconstrained settings.

%% file: sec/6_appendix.tex
\section{Accelerating Training}\label{app:train}
The pose estimator estimates human pose from IMU measurements and the gravity direction in the root frame.
When using a non-inertial root frame, it is necessary to model the fictitious accelerations induced by fictitious forces.
In previous work PNP~\cite{yi2024pnp}, fictitious accelerations are regressed in an auto-regressive manner using an additional fully connected neural network. However, the auto-regressive approach prevents the use of the highly optimized black-box RNN implementation in CUDNN, which processes the entire sequence at once, resulting in slower training.
In our implementation, we retain the concept of incorporating fictitious accelerations but remove the auto-regressive structure. 
Instead, we combine the fictitious acceleration estimation with the first LSTM, which leverages historical information. 
Specifically, the first LSTM, $PL$, takes the root's local angular velocity and acceleration as additional inputs, which are critical for modeling non-inertial effects of the root coordinate frame. 
The output remains unchanged.
The goal is for the network to automatically learn to model the fictitious accelerations by estimating the leaf joint positions.
This adjustment leads to comparable results with significantly faster training speed.
%

\section{Accelerating Inferring}\label{app:infer}
To enable real-time performance, we accelerate key optimizations in our algorithm.
\paragraph{Root velocity refinement}
In the translation estimator, we use joint stationary constraints to refine the root velocity estimate. The optimization problem in Eq.~\ref{eq:v} can be solved analytically by finding the roots of its derivative. The solution is:
\begin{equation}
    \tilde{\boldsymbol{v}}^t=\frac{1}{1+\sum_is_i}\boldsymbol{v}^t+\sum_i \frac{s_i}{1+\sum_is_i}\frac{1}{\Delta t}\left(\mathrm{FK}_i(\boldsymbol{\theta}^{t-1})-\mathrm{FK}_i(\boldsymbol{\theta}^t)\right).
\end{equation}
\paragraph{Physics-based tracking}
The pre-tracking and re-tracking steps involve solving a quadratic programming problem as presented in Eq.~\ref{eq:pretrack} and \ref{eq:retrack}. However, by substituting the equality constraints into the objective function to eliminate $\boldsymbol{\tau}$, the problem transforms into an unconstrained sparse least squares problem, which can be solved efficiently \hl{using the LSQR method~\cite{paige1982lsqr}}. 
We present the equivalent problem to Eq.~\ref{eq:pretrack} in the sparse least squares formulation:
\begin{equation}\label{eq:acc-pretrack}
    \min_{\ddot{\boldsymbol{q}}}
    \left\|
        \left(\begin{array}{c}
             \boldsymbol{A} \\ \boldsymbol{J} \\ \sqrt{\beta_\tau}\boldsymbol{M}
        \end{array}\right)\ddot{\boldsymbol{q}}-
        \left(\begin{array}{c}
             \ddot{\boldsymbol{\theta}}_{\mathrm{des}} \\ -\dot{\boldsymbol{J}}\dot{\boldsymbol{q}} + \ddot{\boldsymbol{r}}_{\mathrm{des}} \\ -\sqrt{\beta_\tau}\boldsymbol{h}
        \end{array}\right)
    \right\|^2,
\end{equation}
where $\boldsymbol{A}=\left(\begin{array}{cc} \boldsymbol{O} & \boldsymbol{I} \end{array}\right)$ selects the corresponding entries of $\ddot{\boldsymbol{q}}$.
Eq.~\ref{eq:retrack} can be accelerated in a similar manner, with only a slight modification compared to Eq.~\ref{eq:acc-pretrack}:
\begin{equation}\label{eq:acc-retrack}
    \min_{\ddot{\boldsymbol{q}}^*}
    \left\|
        \left(\begin{array}{c}
             \boldsymbol{A} \\ \boldsymbol{J} \\ \sqrt{\beta_\tau^*}\boldsymbol{M}
        \end{array}\right)\ddot{\boldsymbol{q}}^*-
        \left(\begin{array}{c}
             \ddot{\boldsymbol{\theta}}_{\mathrm{des}} \\ -\dot{\boldsymbol{J}}\dot{\boldsymbol{q}} + \ddot{\boldsymbol{r}}_{\mathrm{des}}^* \\ \sqrt{\beta_\tau^*}(-\boldsymbol{h}+\boldsymbol{J}^T\boldsymbol{\lambda})
        \end{array}\right)
    \right\|^2.
\end{equation}
Solving Eq.~\ref{eq:acc-pretrack} and \ref{eq:acc-retrack} yields $\ddot{\boldsymbol{q}}$ and $\ddot{\boldsymbol{q}}^*$, respectively, from which we can compute $\boldsymbol{\tau}$ and $\boldsymbol{\tau}^*$ using Eq.~\ref{eq:tau1} for pre-tracking and Eq.~\ref{eq:tau2} for re-tracking, respectively. \hl{In our implementation, we utilize the LSQR solver from the SciPy library.}